\documentclass[11pt]{article}
\usepackage{color}
\usepackage{subfigure}
\usepackage{amsmath}
\usepackage{amssymb}
\usepackage{graphicx,color}
\usepackage{cite}
\usepackage{enumerate}
\usepackage{amsthm}% theorems and proofs
\usepackage{amsfonts,mathrsfs}
\usepackage{geometry}
\usepackage{ifthen}
\usepackage{graphicx}
\usepackage{epstopdf}
\usepackage{float}
\usepackage{bm}
\usepackage[caption=false]{subfig}
\usepackage[utf8]{inputenc}

\begin{document}

\title{\bf Coherence and Imaginarity as Resources in Quantum Circuit Complexity }

\vskip0.1in
\author{
\begin{minipage}{1.0\textwidth}
\centering
{\small Linlin Ye$^1$, Zhaoqi Wu$^1$\thanks{Corresponding author. E-mail: wuzhaoqi\_conquer@163.com}, Nanrun Zhou$^{2,3}$\thanks{Corresponding author. E-mail: nrzhou@sues.edu.cn}}\\
{\small\it 1. Department of Mathematics, Nanchang University,
Nanchang 330031, P R China}\\
{\small\it 2. School of Electronic and Electrical Engineering, Shanghai University Engineering Science, Shanghai 201620, China}\\
{\small\it 3. Department of Electronic Information Engineering, Nanchang University, Nanchang 330031, China}
\end{minipage}
}

\date{}
\maketitle

\noindent {\bf Abstract} {\small }\\
Quantum circuit complexity quantifies the minimal number of gates
needed to realize a unitary transformation and plays a central role
in quantum computation. In this work, we investigate the complexity
of quantum circuits through coherence and imaginarity resources.
We establish a lower bound on the circuit cost by
the Tsallis relative $\alpha$ entropy of cohering power, which is
shown to be tighter than the one presented by Bu et al.
[\textit{Communications in Mathematical Physics} 405, no. 7 (2024):
161] under restrictive conditions. As a consequence, we obtain the
relationships between the circuit cost and the coherence generating
power via probabilistic average in terms of skew
information/relative entropy, and present explicit
bounds of the circuit cost for typical quantum gates. Moreover, we
derive lower bounds on the circuit cost via the imaginaring power of
the circuit, induced by the Tsallis relative $\alpha$ entropy and
relative entropy. We demonstrate that imaginarity
can yield nontrivial constraints on the circuit cost even when
coherence-based lower bounds are zero (e.g., for the $T$ gate),
which implies that imaginarity may provide advantages under certain
circumstances compared with coherence. Our results may help better
understand the connections between quantum resources and circuit
complexity.


\vskip0.2in

\noindent {\bf 1. Introduction}\\\hspace*{\fill}\\
Originally introduced to explain interference phenomena in light,
quantum coherence is a fundamental feature of quantum
mechanics\cite{SEC,GRJ,SMO}. Moreover, coherence has demonstrated
remarkable utility in interdisciplinary applications, ranging from
quantum thermodynamics~\cite{LM,MLK,VNG,MHJ,PK} and quantum phase
transitions~\cite{JJ} to complex biological systems~\cite{HS,LS},
thereby underscoring its central role in quantum science and
technology. The characterization and quantification
of coherence have attracted widespread attention in recent
years~\cite{TB,YXD,Hu2018,Streltsov2017,WuZZF2021}. Besides
considering coherence with respect to orthonormal bases, coherence
relative to positive operator-valued measures and,
more generally, quantum
measurements~\cite{Bischof2019,Xu2020,Bischof2021}, set
coherence~\cite{Designolle2021}, coherence with respect to
non-orthogonal bases~\cite{Theurer2017,Torun2021}, multi-level
coherence~\cite{Johnston2018,Ringbauer2018,Regula2018,Johnston2022},
have also been investigated. On the other hand,
operational resource-theoretic and channel-resource perspectives
have also been developed \cite{Bu2017MaxRelEntCoherence,
BuXiong2017CoheringPower, LiuBuTakagi2019OneShot,
LiBuLiu2020Channels,TakagiRegulaBuLiuAdesso2019Subchannel}. In
addition, Mani et al. \cite{Mani2015} were the first to introduce
the concept of the cohering power of a generic quantum channel. By
optimizing the output coherence, the coherence generating power
(CGP) of a quantum channel was subsequently defined as a
quantitative measure of its ability to generate quantum coherence.
Building on this foundational work, Zanardi et al.
\cite{Zanardi2017a,Zanardi2017b} employed probabilistic averages to
investigate CGP for the first time. They proposed a method to
quantify the CGP of unitary channels by introducing a measure based
on the average coherence produced by the channel when acting on a
uniform ensemble of incoherent states.

Complex numbers are widely applied in physics, including mechanics,
optics, and electromagnetism. While in classical physics they mainly
simplify models of oscillations and waves, in quantum physics they
play a much deeper
role~\cite{Wootters2012,Hardy2012,Aleksandrova2013}.
Renou et al. \cite{Renou2021} tested an
entanglement-swapping scenario and found that complex-valued quantum
mechanics agrees with the experimental data, whereas real-valued
quantum mechanics shows clear deviations. These results provide
experimental evidence that complex numbers are essential in quantum
mechanics. Subsequent experiments provided direct refutations of
``real-only" quantum mechanics, including superconducting-qubit
implementations that significantly violate the bounds implied by
real quantum theory \cite{ChenM2022} and photonic-network tests
under strict locality and independent-source assumptions that
likewise exclude real-valued quantum mechanics \cite{WuD2022}.
Owing to the special role of imaginary numbers in quantum theory,
Hickey and Gour~\cite{Hickey2018} proposed the resource theory of
imaginarity, based on the imaginary parts of a quantum state's
density matrix, and analyzed pure-state transformations under
single-copy measurements. As the imaginary components of a density
matrix are invariably confined to its off-diagonal elements, the
theory of imaginarity is intrinsically linked to the theory of
coherence. Wu~et~al.~\cite{Wu2021a,Wu2021b} demonstrated both
theoretically and experimentally that complex numbers are crucial in
quantum state discrimination. Unitary-invariant
witnesses of quantum imaginarity\cite{Fernandes2024PRL} and
multistate imaginarity in qubit systems\cite{LiMaosheng2026PRA} have
also been explored and studied. Notably, imaginarity has played
important roles and found wide applications in various fields such
as quantum machine learning\cite{Sajjan2023PRR},
pseudorandom unitaries \cite{Haug2025}, and quantum speed
limit\cite{Xuan2025}.
Moreover, Zhang et al. \cite{Zhang2023} initiated
the study of imaginarity resource theory at the channel level by
first analyzing the imaginaring and deimaginaring powers of qubit
quantum channels.

A fundamental challenge in quantum information and computation is to
determine the complexity of implementing a target unitary $U$,
typically defined as the minimal number of basic gates required to
synthesize it from a fiducial state~\cite{Nielsen,Kitaev}. To
characterize circuit complexity, Nielsen et al. introduced the
related concept of circuit cost in a series of seminal
papers~\cite{Nielsen2006a,Nielsen2006b,Nielsen2006c}. In recent
years, circuit complexity and circuit cost have been shown to play
an important role in high-energy physics \cite{BRSZS,BSSZ,Suss} and
quantum machine learning\cite{GLHCCZ}. Studies have further explored
circuit complexity within the framework of quantum field
theories\cite{Jefferson2017,CHMPF,Bhattacharyya2018,Takayanagi2018},
with particular attention to topological quantum field
theory\cite{Couch2022} and conformal field
theory\cite{Chagnet2022,Bhattacharyya2023}. The
submaximal complexity, termed uncomplexity, serves as a resource for
quantum computation \cite{BrownSusskind2018}, and has since been
formalized within a resource-theoretic framework \cite{Halpern2022};
related resource-theoretic formulations have also been developed for
quantum scrambling\cite{GarciaBuJaffe2023Scrambling}. From a
circuit-theoretic viewpoint, quantum higher-order Fourier analysis
provides an analytic characterization of the Clifford hierarchy
\cite{Bu2025Fourier}, while displaced fermionic Gaussian states
admit efficient classical simulation \cite{LyuBu2025Displaced}.
Eisert demonstrated a clear link between quantum entanglement and
circuit complexity, showing that the entangling power of a unitary
transformation provides a lower bound on its circuit
cost~\cite{Eisert2021}. Furthermore, Bu et al.
established lower bounds on the circuit cost of a quantum circuit by
analyzing its circuit sensitivity, magic power, and cohering
power~\cite{Bu2024}. Li et al. subsequently
introduced a lower bound on quantum circuit complexity based on the
Wasserstein complexity~\cite{Li2025}. Bu et al.
derived bounds on the statistical complexity of quantum circuits by
employing the Rademacher and Gaussian
complexities~\cite{Bu2022Complexity,Bu2023}. In
this work, we establish lower bounds on the circuit cost of quantum
circuits based on the resource rate under Hamiltonian evolution,
following the approach proposed by Bu et al. \cite{Bu2024} in the
study of quantum circuit complexity.
%Quantum coherence has emerged as a critical resource underpinning the efficacy of quantum algorithms \cite{FSH,RPF,MMY,CCH,YWF1,YWF2,YWF3,NMTV,Feng}. Studies have demonstrated that the Deutsch-Jozsa algorithm's precision correlates directly with recoverable coherence\cite{HMC,JMM}. Extensive research has been conducted to elucidate the coherence dynamics in Grover's search algorithm under both ideal and noise-affected conditions, establishing significant correlations between coherence measures and algorithmic success probability \cite{PMQ,MPH,Anand,Shi,Chin,Rastegin1,Rastegin2,Liu,Rastegin3}. Similarly, coherence establishes both upper and lower bounds on Shor's algorithm's performance\cite{AFTTE}. More recently, Zhou showed that the generalized Grover algorithm's success depends not only on oracle queries but also coherence fraction \cite{ZSQ}, confirming coherence as a fundamental determinant of quantum computational advantage.

Building on the conceptual route inherited from
\cite{Bu2024}, in this paper, we further discuss the relationship
between coherence and circuit complexity. We first derive the
explicit expression of the coherence rate based on Tsallis-$\alpha$
relative entropy, which is more general than
relative entropy used in \cite{Bu2024}, and derive the upper bounds
of it. Based on this, utilizing the technique of
Trotter decomposition, we get new lower bounds of the circuit cost
via Tsallis-$\alpha$ relative entropy of cohering power. Letting
$\alpha\rightarrow 1$ and imposing some hypothesis on the input
state, it is found that our bound is tighter than the one in
\cite{Bu2024}. On the other hand, the connection between imaginarity
and circuit cost remains unexplored and poorly understood as far as
we know. In this paper, we fill this gap by
studying this problem. Interestingly, it is found that instead of
coherence, imaginarity may provide nontrivial bounds of circuit cost
for some specific quantum gates, demonstrating the differences of
the two resources.

The remainder of this paper is organized as follows. In Section 2,
we review circuit cost, Tsallis relative $\alpha$ entropy of
coherence, CGP of quantum channels under skew information and
relative entropy. In Section 3, we investigate the relationship
between Tsallis relative $\alpha$ entropy of coherence and circuit
cost. Furthermore, we derive the connections between the circuit
cost and the CGP defined respectively in terms of skew information
and relative entropy. In Section 4, we shift our focus to Tsallis
relative $\alpha$ entropy of imaginarity and relative entropy of
imaginarity, analyze their connections to the circuit cost. Finally,
we summarize the results in Section 5.

\vskip0.2in

\noindent {\bf 2. Circuit cost, coherence and imaginarity}\\\hspace*{\fill}\\
In this section, we review the concepts of circuit cost, Tsallis relative
$\alpha$ entropy of coherence and CGP of quantum channels under skew information of
coherence and relative entropy of coherence. Furthermore, we recall the notion of Tsallis relative $\alpha$ entropy of imaginarity and relative entropy of imaginarity.

Given a fixed gate set, the exact circuit complexity of a target unitary $U$ is commonly defined as the minimum number of quantum gates required to implement $U$ exactly. In practice, one often considers an approximate variant, where it suffices to realize an operation that approximates $U$ within a prescribed and sufficiently small error in the operator norm \cite{Nielsen2006a,Nielsen2006b}.

To connect circuit complexity with a physically
motivated, continuous-time viewpoint, Nielsen et al.\ recast the
synthesis of $U$ as an optimal control problem: $U$ is generated by
a time-dependent Hamiltonian $H(t)$ via the Schr\"odinger equation,
and imposing a cost functional on $H(t)$ induces a notion of path
length, and hence a distance, on the unitary group manifold
\cite{Nielsen2006a}. Under this geometric formulation, searching for
an optimal circuit (equivalently, an optimal control protocol) can
be viewed as finding a shortest path (geodesic) connecting $I$ and
$U$. The resulting minimal distance, often referred to as the
circuit cost, provides for suitable choices of metric, a rigorous
lower bound (up to constant factors) on gate-count complexity and
enables systematic analysis using tools from differential geometry
and the calculus of variations \cite{Nielsen2006a,Nielsen2006b}.
Fig. 1 illustrates the geometric viewpoint: circuit cost equals the
length of the shortest admissible path on $\text{SU}(d^n)$
connecting $I$ and $U$.

Importantly, circuit cost serves as a continuous surrogate for the target circuit complexity: by establishing computable or analytically tractable lower bounds on circuit cost in terms of appropriate resource measures, one can reduce the task of proving circuit lower bounds to estimating these resources and translating them into explicit lower bounds on circuit cost, and consequently on circuit complexity \cite{Eisert2021,Bu2024}.

Let $ U \in \text{SU}(d^n) $ represent a unitary
operator and $ o_1, \dots, o_m $ be normalized traceless Hermitian
operators supported on two qudits with $ \| o_i \|_\infty = 1 $ for
$ i = 1, \dots, m $. The circuit cost of $ U $, with respect to $
o_1, \dots, o_m $, is defined as
\cite{Nielsen2006a,Nielsen2006b,Eisert2021}
\begin{equation}\label{eq1}
\text{Cost}(U) = \inf \int_0^1 \sum_{j=1}^{m} | r_j(s) | \, \mathrm{d}s,
\end{equation}
where $ | r_j(s) | $ represents the absolute value of $ r_j(s) $, and the infimum above is taken over all continuous functions $ r_j : [0, 1] \to \mathbb{R} $ satisfying
\begin{equation}\label{eq2}
U = \mathcal{P} \exp \left( -\mathrm{i} \int_0^1 H(s) \, \mathrm{d}s \right),
\end{equation}
and $H(s) = \sum_{j=1}^{m} r_j(s) o_j$,
where $ \mathcal{P} $ denotes the path-ordering operator.
\begin{figure}[ht]
\centering {\begin{minipage}[figure1]{0.6\linewidth}
\includegraphics[width=0.9\textwidth]{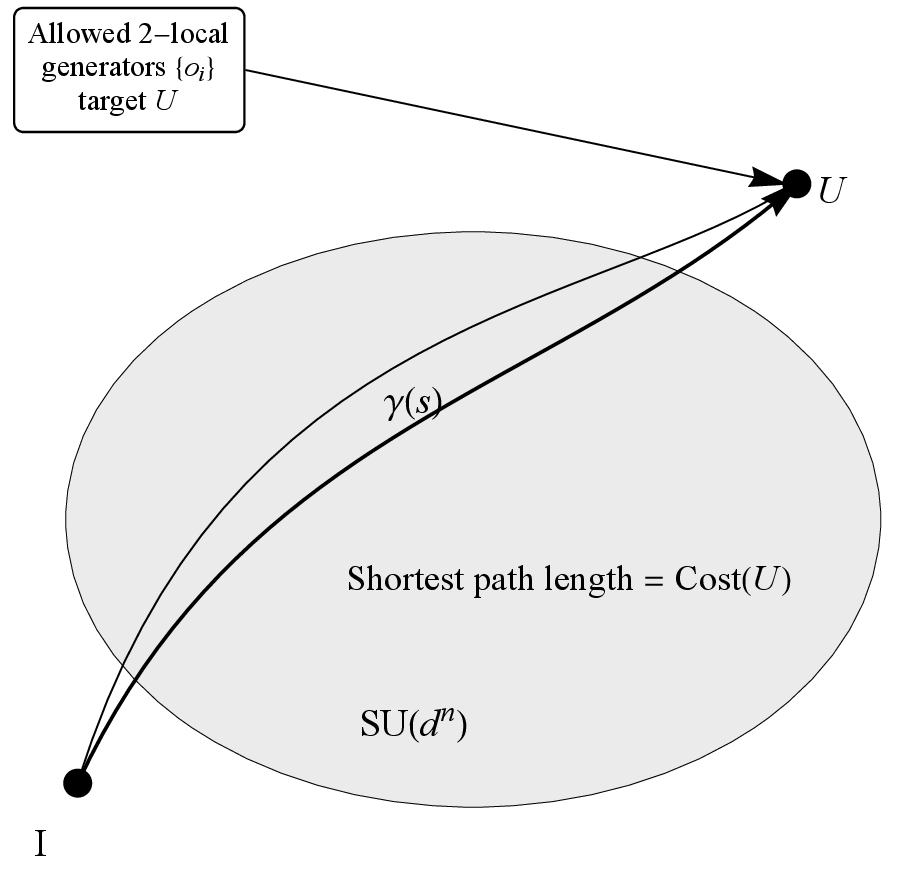}
\end{minipage}}
\caption{{$\mathrm{Cost}(U)$ is the length of the shortest
admissible path from $I$ to $U$ on
$\text{SU}(d^n)$.\label{fig:Fig1}}}
\end{figure}

Denote by $\mathcal{H}$ a $d$ dimensional Hilbert
space, and $\mathcal{D(H)}$ the set of all density operators on
$\mathcal{H}$. The Tsallis relative $\alpha$ entropy provides a
one-parameter generalization of the quantum relative entropy, which
is given by \cite{AS1,AS2}
\begin{equation}\label{eq3}
D_{\alpha}(\rho\|\sigma) = \frac{1}{\alpha - 1} \left[
f_{\alpha}(\rho,\sigma) - 1 \right], \quad \alpha \in (0,1) \cup
(1,+\infty),
\end{equation}
where $f_{\alpha}(\rho,\sigma) =
\mathrm{Tr}\!\left(\rho^{\alpha}\sigma^{1-\alpha}\right)$. When
$\alpha \to 1$, this formula reduces to $S'(\rho\|\sigma) = \ln
2\cdot\, S(\rho\|\sigma)$, with $S(\rho\|\sigma) = \mathrm{Tr}(\rho
\log \rho) - \mathrm{Tr}(\rho \log \sigma)$ denoting the
quantum relative entropy. Throughout the paper, the
logarithm `log' is taken to be base 2. Fixing a reference basis $\{
|j\rangle \}_{j=1}^d$ of $\mathcal{H}$, the Tsallis relative
$\alpha$ entropy of coherence for $\alpha \in (0,1) \cup (1,2]$ is
\cite{ZHYC}
\begin{equation}\label{eq4}
C_{\alpha}(\rho) = \min_{\sigma \in \mathcal{I}} \frac{1}{\alpha - 1}
\left[ f_{\alpha}^{\,1/\alpha}(\rho,\sigma) - 1 \right]
= \frac{1}{\alpha - 1} \left[ \sum_{j=1}^d \langle j | \rho^{\alpha} | j \rangle^{1/\alpha} - 1 \right].
\end{equation}
where $\mathcal{I}$ denotes the set of incoherent states, which are
diagonal in the reference basis. In the limit $\alpha \to 1$,
$C_{\alpha}(\rho)$ reduces to $\ln 2\cdot C_{r}(\rho)$, where
$C_{r}(\rho) = \mathrm{Tr}(\rho\log\rho) -
\mathrm{Tr}(\rho_{\mathrm{diag}} \log $ $\rho_{\mathrm{diag}})$ is
the relative entropy of coherence \cite{TB}. When $\alpha =
\frac12$, $C_{\alpha}(\rho)$ reduces to $2 C_s(\rho)$, with
$C_s(\rho) = 1 - \sum_{j=1}^d \langle j| \sqrt{\rho} | j
\rangle^{2}$ denoting the skew information of coherence\cite{YCS}.

Let $\Phi$ be a quantum channel, namely a completely positive and
trace-preserving (CPTP) map. To quantify its ability to generate
coherence, the method of probabilistic averaging has been
employed~\cite{Zanardi2017a,Zanardi2017b,Zhang2018}. In this
setting, the coherence generating power (CGP) of $\Phi$ is defined
as the average skew information-based coherence produced by the
channel when it acts on a uniformly distributed ensemble of
incoherent states. The CGP of $\Phi$ based on skew information of
coherence and relative entropy of coherence
are\cite{Wu2022,Zhang2018}
\begin{equation}\label{eq5}
\mathrm{CGP}_{S}(\Phi) := \int_{\mathcal{I}} \mathrm{d}\mu(\rho)\, C_{s}\!\left(\Phi(\rho)\right)~~~ \mathrm{and}~~~
\mathrm{CGP}_{R}(\Phi) := \int_{\mathcal{I}} \mathrm{d}\mu(\rho)\, C_{r}\!\left(\Phi(\rho)\right),
\end{equation}
respectively, where $\mathcal{I}$ denotes the set of incoherent
states, $\mu$ refers to the probability measure corresponding to a
uniform ensemble of such states. For incoherent channels
$\Phi_{IO}$, one has $\mathrm{CGP}_{S}(\Phi_{IO}) = 0$ and
$\mathrm{CGP}_{R}(\Phi_{IO}) = 0$, since the output of $\Phi_{IO}$
remains incoherent for all inputs. We consider the special case of
unitary channels. For a unitary operator $U$, the corresponding
channel $\Phi_{U}$ is given by $\Phi_{U}(\rho) = U \rho
U^{\dagger}$, where $U$ is a unitary transformation and
$\dagger$ denotes the Hermitian adjoint.

The imaginarity measure based on Tsallis relative $\alpha$ entropy denoted by $M_{\alpha}(\rho)$,
is defined as\cite{Xu2024}
\begin{equation}\label{eq6}
M_{\alpha}(\rho) = 1 - \mathrm{Tr}\left[\rho^{\alpha}(\rho^{*})^{1-\alpha}\right],
\end{equation}
where $\alpha \in (0,1)$ and $*$ represents (complex) conjugate.
The relative entropy of imaginarity for a quantum state $\rho$ is defined as \cite{Xue2021}
\begin{equation}\label{eq7}
M_r(\rho) = \min_{\sigma \in \mathcal{F}} S(\rho \| \sigma),
\end{equation}
where $S(\rho \| \sigma)= \mathrm{Tr}(\rho\log\rho)-
\mathrm{Tr}(\rho\log\sigma)$ denotes the quantum relative entropy,
and $\mathcal{F}$ denotes the set of real quantum states. Any
quantum state $\rho$ can be decomposed as $\rho = \mathrm{Re}(\rho)
+ \mathrm{i}\,\mathrm{Im}(\rho)$, where $\mathrm{Re}(\rho) =
\frac{1}{2} \big( \rho + \rho^{T} \big)$, $\mathrm{Im}(\rho) =
\frac{1}{2\mathrm{i}} \big( \rho - \rho^{T} \big)$ and $T$
represents the transpose. For any quantum state $\rho$,  the mapping
$\Delta_{1}$ is defined by\cite{ChenG2023}
\begin{equation}\label{8}
\Delta_{1}(\rho) = \frac{1}{2} \left( \rho + \rho^{T} \right).
\end{equation}
Then, the relative entropy of imaginarity $M_{r}(\rho)$ can be
equivalently expressed as\cite{Xue2021}
\begin{equation}\label{9}
M_{r}(\rho) = S\big(\Delta_{1}(\rho)\big) - S(\rho),
\end{equation}
where $S(\rho)=-\mathrm{Tr}(\rho\log\rho)$ is the von Neumann entropy of $\rho$.
\vskip0.2in

\noindent {\bf 3. Coherence and circuit complexity}\\\hspace*{\fill}\\
We now investigate Tsallis relative $\alpha$ entropy of coherence in circuit complexity, and establish a lower bound on the circuit cost based on this coherence. Furthermore, we derive the relationships between the $\mathrm{CGP}_{S}$, $\mathrm{CGP}_{R}$ and the circuit cost of a quantum circuit.

Based on Tsallis relative $\alpha$ entropy of
coherence, we define the cohering power associated with a unitary
evolution $U$ as
\begin{equation}\label{eq10}
\mathcal{C}_\alpha(U) := \max_{\rho \in \mathcal{D}((\mathbb{C}^d)^{\otimes n})} \left| C_\alpha\left(U \rho U^\dagger\right) - C_\alpha(\rho) \right|,
\end{equation}
where the maximization is performed over all density operators
$\rho$. We next introduce the notion of the rate of coherence, which
measures the instantaneous change of a coherence measure when a
state evolves under a Hamiltonian $H$. Under the
Tsallis relative $\alpha$-entropy cohering power, the rate of
coherence is
\begin{equation}\label{eq11}
R^{\alpha}_{C}(H, \rho) := \frac{\mathrm{d}}{\mathrm{d}t} \, C_\alpha \left( e^{-\mathrm{i} t H} \rho \, e^{\mathrm{i} t H} \right) \bigg|_{t=0}.
\end{equation}

The following lemma provides an alternative expression for the rate of coherence in terms of a commutator involving $\rho$ and its dephased version.\\
\textbf{Lemma 1} Given a Hamiltonian $H$ on an $n$ qudit system and
a quantum state $\rho \in \mathcal{D}\big((\mathbb{C}^d)^{\otimes
n}\big)$, for $\alpha \in (0,1)$, the rate of
coherence can be expressed as
\begin{equation}\label{eq12}
R^{\alpha}_{C}(H, \rho) = \frac{\mathrm{i}}{\alpha(1-\alpha)} \, \mathrm{Tr}\left( \left[ \rho^{\alpha}, (\Delta(\rho^{\alpha}))^{\frac{1}{\alpha}-1} \right] H\right),
\end{equation}
where $\Delta(\cdot)=\sum_{i} |i\rangle \langle i| \, \cdot \,
|i\rangle \langle i|$ denotes the completely dephasing channel.
For $\alpha \in (1,2]$, Eq. (\ref{eq12}) also
holds, if $\Delta(\rho^\alpha)$ is strictly positive in the
reference basis, or equivalently, $p_j=
\langle j|\rho^\alpha|j\rangle>0$ for all $j$.\\
$\it{Proof}$. Let $\rho_t = e^{-\mathrm{i} t H} \rho \, e^{\mathrm{i} t H}$ denote the state at time $t$.
The coherence rate can be written as
\begin{equation*}
R^{\alpha}_{C}(H, \rho) = \frac{\mathrm{d}}{\mathrm{d}t} C_{\alpha}(\rho_t) \bigg|_{t=0}
=\frac{1}{\alpha - 1} \frac{\mathrm{d}}{\mathrm{d}t} \left[ \sum_{j=1}^d \langle j | \rho_t^{\alpha} | j \rangle^{1/\alpha} - 1 \right]\bigg|_{t=0}.
\end{equation*}
Differentiating $\rho_t^{\alpha}$ with respect to $t$ and evaluating at $t=0$, we obtain
\begin{equation*}
\frac{\mathrm{d}}{\mathrm{d}t} \rho_t^{\alpha}\bigg|_{t=0} = \frac{\mathrm{d}}{\mathrm{d}t} \left( e^{-\mathrm{i} H t} \, \rho \, e^{\mathrm{i} H t} \right)^{\alpha}\bigg|_{t=0}= \frac{\mathrm{d}}{\mathrm{d}t}\left( e^{-\mathrm{i} H t} \, \rho^{\alpha} \, e^{\mathrm{i} H t} \right)\bigg|_{t=0}=-\mathrm{i}H\rho^\alpha+\mathrm{i}\rho^\alpha H=\mathrm{i}\left[\rho^{\alpha},H\right].
\end{equation*}
Thus, by the definition of the trace, we have
\begin{align*}
R^{\alpha}_{C}(H, \rho)
&=\frac{\mathrm{i}}{(\alpha - 1)\alpha} \sum_j \langle j | \rho^{\alpha} | j \rangle^{\frac{1}{\alpha} - 1} \langle j | [\rho^{\alpha}, H] | j \rangle\notag\\
&=\frac{\mathrm{i}}{(\alpha - 1)\alpha} \sum_j \langle j | \Delta(\rho^{\alpha}) | j \rangle^{\frac{1}{\alpha} - 1} \langle j | [\rho^{\alpha}, H] | j \rangle\notag\\
&=\frac{\mathrm{i}}{(\alpha - 1)\alpha}\mathrm{Tr}\left((\Delta(\rho^{\alpha}))^{\frac{1}{\alpha}-1} \left[\rho^{\alpha}, H\right]\right)\notag\\
&= \frac{\mathrm{i}}{\alpha(1-\alpha)} \, \mathrm{Tr}\left( \left[
\rho^{\alpha}, (\Delta(\rho^{\alpha}))^{\frac{1}{\alpha}-1} \right]
H\right).
\end{align*}
Note that for $\alpha\in(1,2]$, $\langle j |
\rho^{\alpha} | j \rangle^{\frac{1}{\alpha} - 1}$ is well defined
since $p_j= \langle j|\rho^\alpha|j\rangle>0$. This completes the
proof.\qed

In particular, in the limit $\alpha \to 1$, $C_{\alpha}(\rho)$ converges to $\ln 2 \cdot C_{r}(\rho)$. Consequently, the resulting expression coincides with the coherence rate reported in \cite{Bu2024}.
Next, we study the upper bound of the coherence rate.\\
{\bf Lemma 2} Given a Hamiltonian $H$ on an $n$ qudit system and an $n$ qudit quantum state
$\rho \in \mathcal{D}\big((\mathbb{C}^d)^{\otimes n}\big)$,
the coherence rate satisfies the following bound.\\
(1) When $\alpha\in (0,1)$, we obtain
\begin{equation}\label{eq13}
\big| R^{\alpha}_{C}(H,\rho) \big| \le
\frac{1}{\alpha(1-\alpha)}\,\|H\|_{\infty}\,\mathrm{Tr}(\rho^{\alpha});
\end{equation}
(2) When $\alpha\in (1,2]$, assume that $\Delta(\rho^\alpha)$ is
strictly positive in the reference basis (equivalently $p_j =
\langle j | \rho^\alpha | j \rangle > 0$ for all $j$). Then we have
\begin{equation}\label{eq14}
\big| R^{\alpha}_{C}(H,\rho) \big| \le\frac{1}{\alpha(\alpha-1)}\,\|H\|_{\infty}\,\mathrm{Tr}(\rho^{\alpha})\,p_{\min}^{\frac{1}{\alpha}-1},
\end{equation}
where $p_{\min}=\min_j p_j>0$.\\
$\it{Proof}$. From Lemma 1 and H\"older's
inequality, we obtain
\begin{align*}
|R^{\alpha}_{C}(H, \rho)|
&=\left|\frac{1}{\alpha(1-\alpha)} \, \mathrm{Tr}\left( \left[ \rho^{\alpha}, (\Delta(\rho^{\alpha}))^{\frac{1}{\alpha}-1} \right] H\right)\right|\notag\\
&=\left|\frac{1}{\alpha(1-\alpha)} \right| \mathrm{Tr}\!\left(\rho^{\alpha}\bigl[(\Delta(\rho^{\alpha}))^{\frac{1}{\alpha}-1},H\bigr]\right)\\
&\le \left|\frac{1}{\alpha(1-\alpha)}\right|
\|\rho^{\alpha}\|_{1}\,\left\|\bigl[(\Delta(\rho^{\alpha}))^{\frac{1}{\alpha}-1},H\bigr]\right\|_{\infty}\\
&=\left|\frac{1}{\alpha(1-\alpha)}\right|
\mathrm{Tr}(\rho^\alpha)\,
\left\|\bigl[(\Delta(\rho^{\alpha}))^{\frac{1}{\alpha}-1},H\bigr]\right\|_{\infty}.
\end{align*}
We next estimate
$\left\|\bigl[(\Delta(\rho^{\alpha}))^{\frac{1}{\alpha}-1},H\bigr]\right\|_{\infty}$.
For any constant $c\in\mathbb{R}$, we have
$\bigl[(\Delta(\rho^{\alpha}))^{\frac{1}{\alpha}-1},H\bigr]$
$=\bigl[(\Delta(\rho^{\alpha}))^{\frac{1}{\alpha}-1}-cI,\,H\bigr].$
Moreover, for any operator $X$,
$\|[X,H]\|_{\infty} \le 2\|X\|_{\infty}\,\|H\|_{\infty}.$
Here, we choose $c$ to shift the operator to its spectral center, namely
\begin{equation*}
c=
\frac{
\lambda_{\max}\!\left((\Delta(\rho^{\alpha}))^{\frac{1}{\alpha}-1}\right)
+
\lambda_{\min}\!\left((\Delta(\rho^{\alpha}))^{\frac{1}{\alpha}-1}\right)
}{2},
\end{equation*}
where $\lambda_{\max}\!\left((\Delta(\rho^{\alpha}))^{\frac{1}{\alpha}-1}\right)$ and
$\lambda_{\min}\!\left((\Delta(\rho^{\alpha}))^{\frac{1}{\alpha}-1}\right)$ are the largest and smallest eigenvalues of
$(\Delta(\rho^{\alpha}))^{\frac{1}{\alpha}-1}$, respectively.
With this choice,
\begin{equation*}
\left\|(\Delta(\rho^{\alpha}))^{\frac{1}{\alpha}-1}-cI\right\|_{\infty}
=
\frac{
\lambda_{\max}\!\left((\Delta(\rho^{\alpha}))^{\frac{1}{\alpha}-1}\right)
-
\lambda_{\min}\!\left((\Delta(\rho^{\alpha}))^{\frac{1}{\alpha}-1}\right)
}{2}.
\end{equation*}
Therefore,
\begin{equation*}
\left\|\bigl[(\Delta(\rho^{\alpha}))^{\frac{1}{\alpha}-1},H\bigr]\right\|_{\infty}
\le
\Bigl(
\lambda_{\max}\!\left((\Delta(\rho^{\alpha}))^{\frac{1}{\alpha}-1}\right)
-
\lambda_{\min}\!\left((\Delta(\rho^{\alpha}))^{\frac{1}{\alpha}-1}\right)
\Bigr)\|H\|_{\infty}.
\end{equation*}
For $\alpha\in(1,2]$, we have $\frac{1}{\alpha}-1<0$. Since $p_j>0$
for all $j$, it follows that $p_j^{\frac{1}{\alpha}-1}$ and
$(\Delta(\rho^\alpha))^{\frac{1}{\alpha}-1}$ are well defined and
bounded. Let $p_{\max}=\max_{j} p_j$ and $p_{\min}=\min_{j} p_j$.
Since $(\Delta(\rho^{\alpha}))^{\frac{1}{\alpha}-1}$ is diagonal,
its eigenvalues are $p_j^{\frac{1}{\alpha}-1}$. Hence,
\begin{equation*}
\lambda_{\max}\!\left((\Delta(\rho^{\alpha}))^{\frac{1}{\alpha}-1}\right)
-\lambda_{\min}\!\left((\Delta(\rho^{\alpha}))^{\frac{1}{\alpha}-1}\right)
=
\left|p_{\max}^{\frac{1}{\alpha}-1}
-p_{\min}^{\frac{1}{\alpha}-1}\right|,
\end{equation*}
where $\alpha \in (0,1) \cup (1,2]$.
Combining the previous estimates, we arrive at
\begin{equation*}
\bigl|R^{\alpha}_{C}(H,\rho)\bigr|
\le \left|\frac{1}{\alpha(1-\alpha)}\right|
\mathrm{Tr}(\rho^{\alpha})
\left|p_{\max}^{\frac{1}{\alpha}-1}
-p_{\min}^{\frac{1}{\alpha}-1}\right|\|H\|_{\infty}.
\end{equation*}
If $\alpha\in (0,1)$, it follows that $0\le
p_j^{\frac{1}{\alpha}-1}\le 1$, so
$\left|p_{\max}^{\frac{1}{\alpha}-1}
-p_{\min}^{\frac{1}{\alpha}-1}\right|$ is at most $1$. Therefore,
\begin{equation*}
\bigl|R^{\alpha}_{C}(H,\rho)\bigr|
\le \frac{1}{\alpha(1-\alpha)}\,\|H\|_{\infty}\,\mathrm{Tr}(\rho^{\alpha}).
\end{equation*}
If $\alpha\in (1,2]$, we get $\left|p_{\max}^{\frac{1}{\alpha}-1}
-p_{\min}^{\frac{1}{\alpha}-1}\right|=p_{\min}^{\frac{1}{\alpha}-1}-p_{\max}^{\frac{1}{\alpha}-1}
\le p_{\min}^{\frac{1}{\alpha}-1}$, and hence
\begin{equation*}
\bigl|R^{\alpha}_{C}(H,\rho)\bigr|
\le \frac{1}{\alpha(\alpha-1)}\,\|H\|_{\infty}\,\mathrm{Tr}(\rho^{\alpha})\,p_{\min}^{\frac{1}{\alpha}-1}.
\end{equation*}
The proof is complete.\qed\\
{\bf Remark 1} By letting
$\alpha\to 1$ in the proof of Lemma 2, we obtain
\begin{equation}
\bigl|R_{C_{r}}(H,\rho)\bigr| \le \frac{1}{\ln
2}\ln\frac{p_{\max}}{p_{\min}}\cdot\|H\|_{\infty},
\end{equation}
where $R_{{C_{r}}}(H,\rho)$ is the coherence rate based on relative entropy.\\
{\bf Theorem 1} For an $n$ qudit system with Hamiltonian $H$ acting on a $k$-qudit subsystem, and an $n$ qudit quantum state $\rho \in \mathcal{D}((\mathbb{C}^d)^{\otimes n})$, the following bounds hold.\\
(1) When $\alpha \in (0, 1)$, we have
\begin{equation}\label{eq16}
|R^{\alpha}_{C}(H, \rho)|\le
   \frac{1}{\alpha(1-\alpha)}d^{k(\frac{1}{\alpha}-1)}\|H\|_{\infty};
\end{equation}
(2) When $\alpha\in (1,2]$, assume that $\Delta(\rho^\alpha)$ is
strictly positive in the reference basis (equivalently $p_j =
\langle j | \rho^\alpha | j \rangle > 0$ for all $j$). Then we have
\begin{equation}\label{eq17}
\left|R^{\alpha}_{C}(H, \rho)\right|
\leq \frac{2}{\alpha(\alpha-1)}
\,d^{n\left(1-\frac{1}{\alpha}\right)}
\,\|H\|_{\infty}.
\end{equation}
$\it{Proof}$. Given that $H$ acts on a $k$-qudit subsystem, there exists a subset $S \subset [n]$ with $|S| = k$, such that $H$ can be decomposed as $H = H_S \otimes I_{S^c}$.
Based on Lemma 1, we can decompose the state $|\vec{z}\rangle$ as $|\vec{z}\rangle = |\vec{x}\rangle |\vec{y}\rangle$, where $\vec{x} \in [d]^S$ and $\vec{y} \in [d]^{S^c}$. This decomposition leads to the following expression
\begin{align*}
R^{\alpha}_{C}(H, \rho)
=&\frac{\mathrm{i}}{(1-\alpha)\alpha}\mathrm{Tr}\left( [H,\rho^{\alpha}](\Delta(\rho^{\alpha}))^{\frac{1}{\alpha}-1}\right)\notag\\
=&\frac{\mathrm{i}}{(1-\alpha)\alpha}\sum_{\vec{z} \in [d]^n} \langle \vec{z} | [H_S \otimes I_{S^{c}}, \rho^{\alpha}] | \vec{z} \rangle \langle \vec{z} |(\Delta(\rho^{\alpha}))^{\frac{1}{\alpha}-1}| \vec{z} \rangle\notag\\
=&\frac{\mathrm{i}}{(1-\alpha)\alpha}\sum_{\vec{x} \in [d]^S, \vec{y} \in [d]^{S^{c}}} \langle \vec{x} | \langle\vec{y}| [H_S \otimes I_{S^{c}}, \rho^{\alpha}] |\vec{x} \rangle|\vec{y} \rangle \left(\langle \vec{x} | \langle\vec{y}|\rho^{\alpha}|\vec{x} \rangle|\vec{y} \rangle\right)^{\frac{1}{\alpha}-1}.
\end{align*}
We now define a family of $k$-qudit states $\{\rho_{\vec{y},\alpha}\}_{\vec{y}}$ as follows.
For any $\vec{y} \in [d]^{S^{c}}$ with $p_{\vec{y},\alpha}>0$, define
\begin{equation*}
\rho_{\vec{y},\alpha}
=
\frac{\operatorname{Tr}_{S^{c}}\!\left[\rho^{\alpha}\bigl(|\vec{y}\rangle \langle \vec{y}| \otimes I_S\bigr)\right]}{p_{\vec{y},\alpha}},
\end{equation*}
where the nonnegative weight $p_{\vec{y},\alpha}$ is given by
\begin{equation*}
p_{\vec{y},\alpha}
=
\operatorname{Tr}\!\left[\rho^{\alpha}\bigl(|\vec{y}\rangle \langle \vec{y}| \otimes I_S\bigr)\right].
\end{equation*}
Note that $\sum_{\vec{y}} p_{\vec{y},\alpha} = \operatorname{Tr}(\rho^\alpha)$,
and hence the collection $\{p_{\vec{y},\alpha}\}_{\vec{y}}$ is generally not normalized.
Indices with $p_{\vec{y},\alpha}=0$ are omitted from all subsequent sums.
From this, we have
\begin{equation*}
R^{\alpha}_{C}(H, \rho)= \frac{\mathrm{i}}{(1-\alpha)\alpha}\sum_{\vec{x} \in [d]^S, \vec{y} \in [d]^{S^{c}}}p_{\vec{y},\alpha}^{\frac{1}{\alpha}} \langle \vec{x} | [H_S \otimes I_{S^{c}}, \rho_{\vec{y},\alpha}] |\vec{x}\rangle\langle \vec{x} |\rho_{\vec{y},\alpha}|\vec{x} \rangle^{\frac{1}{\alpha}-1}.
\end{equation*}
Let
$c_{\vec{y},\alpha}= \mathrm{Tr}\!\left(\rho_{\vec{y},\alpha}^{1/\alpha}\right)$ and
$\tilde{\rho}_{\vec{y},\alpha}= \frac{1}{c_{\vec{y},\alpha}}\rho_{\vec{y},\alpha}^{1/\alpha}$.
Similarly, one can see that
\begin{equation*}
\,c_{\vec{y},\alpha}\, R_{\mathcal{C}}^{\alpha}\!\left(H_S,\tilde{\rho}_{\vec{y},\alpha}\right)= \frac{\mathrm{i}}{(1-\alpha)\alpha}\sum_{\vec{x} \in [d]^S}\langle \vec{x} | [H_S, \rho_{\vec{y},\alpha}] |\vec{x}\rangle\langle \vec{x} |\rho_{\vec{y},\alpha}|\vec{x} \rangle^{\frac{1}{\alpha}-1}.
\end{equation*}
Consequently, we obtain
\begin{equation*}
R^{\alpha}_{C}(H, \rho)=\sum_{\vec{y} \in [d]^{S^{c}}}p_{\vec{y},\alpha}^{\frac{1}{\alpha}}\,c_{\vec{y},\alpha}\, R_{\mathcal{C}}^{\alpha}\!\left(H_S,\tilde{\rho}_{\vec{y},\alpha}\right).
\end{equation*}
{\bf Case 1.} $\alpha\in (0,1)$. Define $\tau =
\mathrm{Tr}_S(\rho^\alpha)$, which is a positive semidefinite
operator on $S^c$. In the orthonormal basis $\{|\vec{y}\rangle\}$ of
$S^c$, we simply have $p_{\vec{y},\alpha}= \langle
\vec{y}|\tau|\vec{y} \rangle=\tau_{\vec{y}\vec{y}}$. Since
$\frac{1}{\alpha}>1$, the function $f(t)=t^{\frac{1}{\alpha}}$ is
convex on $[0,+\infty)$. By the Schur-Horn theorem, the diagonal
vector of $\tau$ is majorized by its eigenvalue vector. Hence, by
Karamata's inequality, we obtain
\begin{equation*}
\sum_{\vec{y}} p_{\vec{y},\alpha}^{1/\alpha}=\sum_{\vec{y}} \tau_{\vec{y}\vec{y}}^{1/\alpha}
\le \sum_{j} \lambda_j(\tau)^{1/\alpha}
= \mathrm{Tr}(\tau^{1/\alpha}).
\end{equation*}
For any positive semidefinite operator $\tau$, we have
$\mathrm{Tr}(\tau^{1/\alpha}) = \|\tau\|_{1/\alpha}^{1/\alpha}$. By
Schatten norm duality and H\"older's inequality,
\begin{equation*}
\begin{aligned}
\|\mathrm{Tr}_S (\rho^{\alpha})\|_{1/\alpha}
&= \sup_{\|Y\|_{\frac{1}{1-\alpha}}=1}\, \bigl|\mathrm{Tr}\bigl(Y\,\mathrm{Tr}_S(\rho^{\alpha})\bigr)\bigr|
= \sup_{\|Y\|_{\frac{1}{1-\alpha}}=1}\, \bigl|\mathrm{Tr}\bigl((I_S\otimes Y)\,\rho^{\alpha}\bigr)\bigr| \\
&\le \sup_{\|Y\|_{\frac{1}{1-\alpha}}=1}\, \|I_S\otimes Y\|_{\frac{1}{1-\alpha}}\,\|\rho^{\alpha}\|_{1/\alpha}
= \|I_S\|_{\frac{1}{1-\alpha}}\,\|\rho^{\alpha}\|_{1/\alpha}
= (d^{k})^{1-\alpha}\,\|\rho^{\alpha}\|_{1/\alpha}.
\end{aligned}
\end{equation*}
Moreover, since $\rho$ is a density operator satisfying $\mathrm{Tr}(\rho)=1$, we obtain
\begin{equation*}
\|\rho^{\alpha}\|_{1/\alpha}
= \left(\mathrm{Tr}((\rho^{\alpha})^{1/\alpha})\right)^{\alpha}
= \left(\mathrm{Tr}(\rho)\right)^{\alpha}
= 1.
\end{equation*}
Hence,
$\|\tau\|_{1/\alpha} \le (d^{k})^{1-\alpha}$.
Raising both sides to the power $1/\alpha$ gives
\begin{equation*}
\mathrm{Tr}(\tau^{1/\alpha}) \le (d^{k})^{\frac{1-\alpha}{\alpha}} =
d^{k(\frac{1}{\alpha}-1)}.
\end{equation*}
Finally, we conclude that
\begin{equation*}
\sum_{\vec{y}} p_{\vec{y},\alpha}^{1/\alpha} \le
d^{k(\frac{1}{\alpha}-1)}.
\end{equation*}
From Lemma 2, we have
\begin{equation*}
\left|\,c_{\vec{y},\alpha}\, R_{\mathcal{C}}^{\alpha}\!\left(H_S,\tilde{\rho}_{\vec{y},\alpha}\right)\right|
\le
\frac{1}{\alpha(1-\alpha)}\,\|H_S\|_{\infty}\, c_{\vec{y},\alpha}^{\,1-\alpha}
\le
   \frac{1}{\alpha(1-\alpha)}\|H_S\|_{\infty}.
\end{equation*}
Therefore,
\begin{equation*}
\left|R^{\alpha}_{C}(H, \rho)\right|=\left|\sum_{\vec{y} \in [d]^{S^{c}}}p_{\vec{y},\alpha}^{\frac{1}{\alpha}}c_{\vec{y},\alpha}
R_{\mathcal{C}}^{\alpha}\!\left(H_S,\tilde{\rho}_{\vec{y},\alpha}\right)\right|\le
   \frac{1}{\alpha(1-\alpha)}d^{k(\frac{1}{\alpha}-1)}\|H\|_{\infty}.
\end{equation*}
{\bf Case 2.} $\alpha\in(1,2]$. In this case, the function
$f(t)=t^{\frac{1}{\alpha}}$ is concave on $[0,+\infty)$. By Jensen's
inequality, we obtain
\begin{equation*}
\frac{1}{d^{\,n-k}} \sum_{\vec y} p_{\vec y,\alpha}^{1/\alpha}
\le
\left( \frac{1}{d^{\,n-k}} \sum_{\vec y} p_{\vec y,\alpha} \right)^{1/\alpha}.
\end{equation*}
Multiplying both sides by $d^{\,n-k}$ yields
$\sum_{\vec y} p_{\vec y,\alpha}^{1/\alpha}
\le
d^{(n-k)(1-\frac{1}{\alpha})}
\left( \sum_{\vec y} p_{\vec y,\alpha} \right)^{1/\alpha}.$
Note that $\sum_{\vec y} p_{\vec y,\alpha}
= \mathrm{Tr}(\rho^{\alpha})$. Moreover, since $\rho$ is a density operator and
$\alpha>1$, it follows that $\mathrm{Tr}(\rho^{\alpha}) \le 1$.
Consequently, we have
\begin{equation*}
\sum_{\vec y} p_{\vec y,\alpha}^{1/\alpha}
\le
d^{(n-k)\left(1-\frac{1}{\alpha}\right)}.
\end{equation*}
$\rho_{\vec{y},\alpha}$ is a density operator on the subsystem $S$.
Letting $q_{\vec{x}} = \langle \vec{x} | \rho_{\vec{y},\alpha} |
\vec{x} \rangle$, we get $\sum_{\vec{x}}  q_{\vec{x}}  = 1$. Note
that here $\Delta(\rho_{\vec{y},\alpha})^{\frac{1}{\alpha}-1}$ is
defined on the support of $\Delta(\rho_{\vec{y},\alpha})$
(equivalently, via the Moore-Penrose generalized inverse), so that
the subsequent bounds remain well defined. From Lemma 1, we have
\begin{equation*}
\,c_{\vec{y},\alpha}\, R_{\mathcal{C}}^{\alpha}\!\left(H_S,\tilde{\rho}_{\vec{y},\alpha}\right)
=
\frac{\mathrm{i}}{\alpha(1-\alpha)}
\mathrm{Tr}\!\left([H_S,\rho_{\vec{y},\alpha}]
\,\Delta(\rho_{\vec{y},\alpha})^{\frac{1}{\alpha}-1}\right).
\end{equation*}
Taking absolute values and applying the triangle inequality yields
\begin{equation*}
\left|
\mathrm{Tr}\!\left([H_S,\rho_{\vec{y},\alpha}]
\,\Delta(\rho_{\vec{y},\alpha})^{\frac{1}{\alpha}-1}\right)
\right|
\le
\left|\mathrm{Tr}\!\left(H_S\rho_{\vec{y},\alpha}
\,\Delta(\rho_{\vec{y},\alpha})^{\frac{1}{\alpha}-1}\right)\right|
+
\left|\mathrm{Tr}\!\left(\rho_{\vec{y},\alpha}H_S
\,\Delta(\rho_{\vec{y},\alpha})^{\frac{1}{\alpha}-1}\right)\right|.
\end{equation*}
For the first term, H\"older's inequality implies
\begin{equation*}
\left|\mathrm{Tr}\!\left(H_S\rho_{\vec{y},\alpha}
\,\Delta(\rho_{\vec{y},\alpha})^{\frac{1}{\alpha}-1}\right)\right|
\le
\|H_S\|_{\infty}
\left\|
\rho_{\vec{y},\alpha}
\,\Delta(\rho_{\vec{y},\alpha})^{\frac{1}{\alpha}-1}
\right\|_1.
\end{equation*}
We first assume that $\alpha\in (1,2)$. Applying the Schatten norm
H\"older's inequality yields
\begin{equation*}
\left\|
\rho_{\vec{y},\alpha}
\,\Delta(\rho_{\vec{y},\alpha})^{\frac{1}{\alpha}-1}
\right\|_1
\le
\|\rho_{\vec{y},\alpha}^{1/2}\|_2
\left\|
\rho_{\vec{y},\alpha}^{1/2}
\,\Delta(\rho_{\vec{y},\alpha})^{\frac{1}{\alpha}-1}
\right\|_2.
\end{equation*}
Since $\|\rho_{\vec{y},\alpha}^{1/2}\|_2^2=\mathrm{Tr}(\rho_{\vec{y},\alpha})=1$,
we have $\|\rho_{\vec{y},\alpha}^{1/2}\|_2=1$. Moreover, since $\Delta(\rho_{\vec{y},\alpha})$ is diagonal, we obtain
\begin{equation*}
\left\|
\rho_{\vec{y},\alpha}^{1/2}
\,\Delta(\rho_{\vec{y},\alpha})^{\frac{1}{\alpha}-1}
\right\|_2^2
=
\mathrm{Tr}\!\left(
\rho_{\vec{y},\alpha}
\,\Delta(\rho_{\vec{y},\alpha})^{\frac{2}{\alpha}-2}
\right)=
\sum_{\vec{x}} q_{\vec{x}}^{\frac{2}{\alpha}-1}.
\end{equation*}
Since the function $f(t)=t^{\frac{2}{\alpha}-1}$ is concave on
$[0,+\infty)$ when $\alpha\in (1,2)$, Jensen's inequality on a space
of dimension $d^k$ implies $\sum_{\vec{x}}
q_{\vec{x}}^{\frac{2}{\alpha}-1} \le
(d^k)^{1-\left(\frac{2}{\alpha}-1\right)}$. Consequently, we obtain
\begin{equation*}
\left\|
\rho_{\vec{y},\alpha}^{1/2}
\,\Delta(\rho_{\vec{y},\alpha})^{\frac{1}{\alpha}-1}
\right\|_2
\le (d^k)^{\frac{1}{2}\left(1-\left(\frac{2}{\alpha}-1\right)\right)}
= (d^k)^{1-\frac{1}{\alpha}}.
\end{equation*}
This yields the trace-norm bound
\begin{equation*}
\left\|
\rho_{\vec{y},\alpha}
\,\Delta(\rho_{\vec{y},\alpha})^{\frac{1}{\alpha}-1}
\right\|_1
\le (d^k)^{1-\frac{1}{\alpha}}.
\end{equation*}
By the same argument, we also have
$\left\|
\Delta(\rho_{\vec{y},\alpha})^{\frac{1}{\alpha}-1}
\,\rho_{\vec{y},\alpha}
\right\|_1
\le (d^k)^{1-\frac{1}{\alpha}}$.
Combining these estimates, we obtain
\begin{equation*}
\left|
\mathrm{Tr}\!\left([H_S,\rho_{\vec{y},\alpha}]
\,\Delta(\rho_{\vec{y},\alpha})^{\frac{1}{\alpha}-1}\right)
\right|
\le
2\,\|H_S\|_{\infty}\,(d^k)^{1-\frac{1}{\alpha}}.
\end{equation*}
Finally, inserting this bound into the definition of $R_C^{\alpha}$ yields
\begin{equation*}
\left|
\,c_{\vec{y},\alpha}\, R_{\mathcal{C}}^{\alpha}\!\left(H_S,\tilde{\rho}_{\vec{y},\alpha}\right)
\right|
\le
\frac{2}{\alpha(\alpha-1)}\,
\|H_S\|_{\infty}\,(d^k)^{1-\frac{1}{\alpha}}.
\end{equation*}
It remains to treat the endpoint $\alpha=2$. Under the Moore-Penrose
convention,
\[
\left\|\rho_{\vec y,2}^{1/2}\,\Delta(\rho_{\vec y,2})^{-1/2}\right\|_2^2
= \mathrm{Tr}\!\left(\rho_{\vec y,2}\,\Delta(\rho_{\vec y,2})^{-1}\right)
= \mathrm{rank}\!\big(\Delta(\rho_{\vec y,2})\big)
\le d^k.
\]
Therefore,
\[
\left\|\rho_{\vec y,2}\,\Delta(\rho_{\vec y,2})^{-1/2}\right\|_1
\le \left\|\rho_{\vec y,2}^{1/2}\right\|_2
\left\|\rho_{\vec y,2}^{1/2}\,\Delta(\rho_{\vec y,2})^{-1/2}\right\|_2
\le (d^k)^{1/2},
\]
and the same bound as above follows for $\alpha=2$. Therefore, for
$\alpha\in (1,2]$, we have
\begin{equation*}
\left|R^{\alpha}_{C}(H, \rho)\right|=\left|\sum_{\vec{y} \in [d]^{S^{c}}}p_{\vec{y},\alpha}^{\frac{1}{\alpha}}c_{\vec{y},\alpha}
R_{\mathcal{C}}^{\alpha}\!\left(H_S,\tilde{\rho}_{\vec{y},\alpha}\right)\right|\le
\frac{2}{\alpha(\alpha-1)}
\,d^{n\left(1-\frac{1}{\alpha}\right)}
\,\|H\|_{\infty},
\end{equation*}
which completes the proof. \qed\\
{\bf Remark 2} (1) When $\rho$ is a pure state, we have
$\sum_{\vec{y}} p_{\vec{y},\alpha} = 1$. For $\alpha\in (0,1)$, it
follows that $p_{\vec{y},\alpha}^{\frac{1}{\alpha}}<
p_{\vec{y},\alpha}$, and $\sum_{\vec{y} \in
[d]^{S^{c}}}p_{\vec{y},\alpha}^{\frac{1}{\alpha}}<1$. Therefore, the
rate of coherence satisfies
\begin{equation}\label{eq18}
\left|R^{\alpha}_{C}(H, \rho)\right| \le
  \begin{cases}
    \frac{1}{\alpha(1-\alpha)} \|H\|_{\infty}, & 0<\alpha< 1,\\[3pt]
     \frac{2}{\alpha(\alpha-1)}d^{n\left(1-\frac{1}{\alpha}\right)}\|H\|_{\infty},
    & 1<\alpha\le 2.
  \end{cases}
\end{equation}
(2) When $\alpha = \frac12$, $C_{\alpha}(\rho)$ reduces to $2
C_s(\rho)$. Therefore, the coherence rate based on skew information
satisfies
\begin{equation}\label{eq19}
|R_{C_{s}}(H, \rho)| \le2d^{k}\|H\|_{\infty}.
\end{equation}
(3) When $\alpha\to 1$, assume that there exists a constant $\delta
\in (0,1)$ such that $p_{\min}=\min_{j} p_j$ satisfies $p_{\min}\geq
\delta$. This assumption implies that
$\ln\frac{p_{\max}}{p_{\min}}\leq\ln\frac{1}{\delta}$. Then the
coherence rate based on relative entropy satisfies
\begin{equation}\label{eq20}
\bigl|R_{C_{r}}(H,\rho)\bigr|
\le \frac{-\ln\delta}{\ln2}\|H\|_{\infty}.
\end{equation}
(4) For $\alpha\in (0,1)$, $\mathrm{Tr}(\rho^{\alpha})$ attains its
maximum at the maximally mixed state $\rho = \mathbf{I}/d^{n}$,
where $\mathbf{I}$ is the identity operator, from which
$\mathrm{Tr}(\rho^{\alpha}) = d^{n(1-\alpha)}$. Then we have
\begin{equation}\label{eq21}
\left|R^{\alpha}_{C}(H, \rho)\right|
\leq \frac{1}{\alpha(1-\alpha)}
\,d^{n(1-\alpha)}
\,\|H\|_{\infty}.
\end{equation}
Moreover, if $n<\frac{k}{\alpha}$, then
$d^{n(1-\alpha)}<d^{k(\frac{1}{\alpha}-1)}$, and in this case the
upper bound in Eq. (\ref{eq21}) is tighter than the one in Eq.
(\ref{eq16}).

Theorem 1 provides the state-independent upper
bounds of the coherence rate. It is evident that when $ \alpha \in
(0, 1) $, the upper bound first decreases and then increases as
$\alpha$ increases. When $\rho$ is a pure state, the upper bounds
are independent of $k$, and specifically, for $ \alpha \in (0, 1)$,
the upper bound depends solely on the parameter $\alpha$ and
$\|H\|_{\infty}$, which is independent of $k$, $d$ and $n$.

Next, we discuss the relationship between the Tsallis relative $ \alpha $ entropy of cohering power and the cost of a quantum circuit.\\
{\bf Theorem 2}
The circuit cost of a quantum circuit $U \in \mathrm{SU}(d^n)$ is lower bounded by the Tsallis relative $\alpha$ entropy of cohering power as\\
(1) For $\alpha \in (0, 1)$, we have
\begin{equation}\label{eq22}
    \mathrm{Cost}(U) \;\geq\;
    d^{2\left(1-\frac{1}{\alpha}\right)} (1-\alpha)\alpha \,
    \mathcal{C}_{\alpha}(U);
\end{equation}
(2) For $\alpha \in (1,2]$, one obtains
\begin{equation}\label{eq23}
    \mathrm{Cost}(U) \;\geq\;
    \tfrac{1}{2} \, d^{n\left(\frac{1}{\alpha}-1\right)}(\alpha-1)\alpha \mathcal{C}_{\alpha}(U).
\end{equation}
$\it{Proof}$. For any arbitrarily small $\varepsilon > 0$, by
applying a Trotter decomposition of $U$, we have $\| U - V_N
\|_{\infty} \leq \varepsilon$, where $V_N$ is defined as $V_N :=
\prod_{t=1}^{N} U_t$, with each $U_t$ given by $U_t := \exp\left(
-\frac{\mathrm{i}}{N} \sum_{j=1}^{m} r_j\!\left(\frac{t}{N}\right)
o_j \right).$ Define $\rho_0=\rho$ and
$\rho_t=U_t\rho_{t-1}U_t^\dagger$, so $\rho_N=V_N\rho V_N^\dagger$.
By telescoping and the triangle inequality, we obtain
\begin{equation*}
\big|C_\alpha(\rho_N)-C_\alpha(\rho_0)\big|
\le \sum_{t=1}^N \big|C_\alpha(\rho_t)-C_\alpha(\rho_{t-1})\big|.
\end{equation*}
Fix $t$. We further write $U_t=\lim_{l\to\infty}U_t^{(l)}$ with
\begin{equation*}
U_t^{(l)}=\Big(U_{t,1}^{1/l}\cdots U_{t,m}^{1/l}\Big)^l,\qquad
U_{t,j}=\exp\!\left(-\frac{\mathrm{i}}{N}r_j\!\left(\frac{t}{N}\right)o_j\right).
\end{equation*}
It follows from Theorem V.3.3 in \cite{Bhatia1997} that, in finite
dimensional case, the map $\rho \mapsto \rho^\alpha$ is continuous
on the cone of positive semidefinite matrices. Since taking matrix
elements, applying the scalar map $x \mapsto x^{1/\alpha}$, and
finite summation are all continuous, the explicit expression in
Eq.~(\ref{eq4}) implies that $C_\alpha$ is continuous. Let
$\rho_t^{(l)}=U_t^{(l)}\rho_{t-1}(U_t^{(l)})^\dagger$. Since
$U_t^{(l)}\to U_t(\l\to\infty)$ in operator norm, we have the
trace-norm estimate
\begin{equation*}
\|\rho_t^{(l)}-\rho_t\|_1 \le
\|(U_t^{(l)}-U_t)\rho_{t-1}(U_t^{(l)})^\dagger\|_1
+\|U_t\rho_{t-1}((U_t^{(l)})^\dagger-U_t^\dagger)\|_1 \le
2\|U_t^{(l)}-U_t\|_\infty \rightarrow 0,
\end{equation*}
when $l\to\infty$, where we used $\|AXB\|_1\le
\|A\|_\infty\|X\|_1\|B\|_\infty$, $\|\rho_{t-1}\|_1=1$, and
$\|U_t\|_\infty=\|U_t^{(l)}\|_\infty=1$. Hence
$C_\alpha(\rho_t^{(l)}) \to C_\alpha(\rho_t)(\l\to\infty)$, and thus
\[
\big|C_\alpha(\rho_t) - C_\alpha(\rho_{t-1})\big|
= \lim_{l\to\infty} \big|C_\alpha(\rho_t^{(l)}) - C_\alpha(\rho_{t-1})\big|.
\]
Now expand $U_t^{(l)}$ into $lm$ elementary factors
\begin{equation*}
U_t^{(l)}=\prod_{q=1}^{lm} W_{t,q},\qquad
W_{t,(s-1)m+j}:=\exp\!\left(-\frac{\mathrm{i}}{Nl}r_j\!\left(\frac{t}{N}\right)o_j\right),
\end{equation*}
where $s=1,\dots,l$ and $j=1,\dots,m$. Define intermediate states $\sigma_{t,0}^{(l)}=\rho_{t-1}$ and
$\sigma_{t,q}^{(l)}=W_{t,q}\sigma_{t,q-1}^{(l)}W_{t,q}^\dagger$. Then $\sigma_{t,lm}^{(l)}=\rho_t^{(l)}$ and
\begin{equation*}
\big|C_\alpha(\rho_t^{(l)})-C_\alpha(\rho_{t-1})\big|
\le \sum_{q=1}^{lm}\big|C_\alpha(\sigma_{t,q}^{(l)})-C_\alpha(\sigma_{t,q-1}^{(l)})\big|.
\end{equation*}
For a single factor $W=\exp(-\mathrm{i}\tau o_j)$ with
$\tau=\frac{1}{Nl}r_j(\frac{t}{N})$, set
$f(s):=C_\alpha(e^{-\mathrm{i}s o_j}\sigma e^{\mathrm{i}s o_j})$. By
the fundamental theorem of calculus,
\begin{equation*}
\big|f(\tau)-f(0)\big|\le \int_0^{|\tau|} |f'(s)|\,ds,
\qquad f'(s)=R_C^\alpha(o_j,\sigma_s),
\end{equation*}
where $\sigma_s=e^{-\mathrm{i}s o_j}\sigma e^{\mathrm{i}s o_j}$.
Applying Theorem~1 with $k=2$ and $\|o_j\|_\infty=1$ gives
\begin{equation*}
|R_C^\alpha(o_j,\cdot)|
\le
\begin{cases}
\frac{1}{\alpha(1-\alpha)}\,d^{2(\frac{1}{\alpha}-1}), & 0<\alpha<1,\\[0.4em]
\frac{2}{\alpha(\alpha-1)}\,d^{\,n(1-\frac{1}{\alpha})}, &
1<\alpha\le 2,
\end{cases}
\end{equation*}
where for $\alpha\in (1,2]$, we first apply the bound to the
full-rank approximation
$\rho^{(\varepsilon)}=(1-\varepsilon)\rho+\varepsilon
\mathbf{I}/d^n$ (so that the assumption in Theorem~1(2) holds along
the unitary orbit) and then let $\varepsilon\to 0$ by continuity of
$C_\alpha$. Consequently, summing over the $lm$ factors yields
\begin{equation*}
\big|C_\alpha(\rho_t)-C_\alpha(\rho_{t-1})\big|
\le
\begin{cases}
\frac{d^{\frac{2(1-\alpha)}{\alpha}}}{N\alpha(1-\alpha)}\sum_{j=1}^m\left|r_j\!\left(\frac{t}{N}\right)\right|, & 0<\alpha<1,\\[0.7em]
\frac{2d^{\,n(1-1/\alpha)}}{N\alpha(\alpha-1)}\sum_{j=1}^m\left|r_j\!\left(\frac{t}{N}\right)\right|, & 1<\alpha\le 2.
\end{cases}
\end{equation*}
Summing over $t=1,\dots,N$ and letting $N\to\infty$ (hence $V_N\to U$) gives
\begin{equation*}
\big|C_\alpha(U\rho U^\dagger)-C_\alpha(\rho)\big|
\le
\begin{cases}
\frac{d^{\frac{2(1-\alpha)}{\alpha}}}{\alpha(1-\alpha)} \int_0^1\sum_{j=1}^m |r_j(s)|\,ds, & 0<\alpha<1,\\[0.8em]
\frac{2d^{\,n(1-1/\alpha)}}{\alpha(\alpha-1)} \int_0^1\sum_{j=1}^m |r_j(s)|\,ds, & 1<\alpha\le 2.
\end{cases}
\end{equation*}
Taking the infimum over all implementations yields the corresponding bound in terms of $\mathrm{Cost}(U)$, and then taking the maximum over $\rho$ yields
$C_\alpha(U)\le K_\alpha\,\mathrm{Cost}(U)$ with the stated constants. Rearranging gives Eq.~(\ref{eq22}) and Eq.~(\ref{eq23}).
The proof is complete.\qed\\
{\bf Remark 3} (1) When $\rho$ is a pure state, the
circuit cost of a quantum circuit is lower bounded by the Tsallis
relative $\alpha$ entropy of cohering power as
\begin{equation}\label{eq24}
\text{Cost}(U)\geq
  \begin{cases}
   (1-\alpha)\alpha \mathcal{C}_{\alpha}(U), & 0<\alpha< 1,\\[3pt]
    \tfrac{1}{2} \, d^{n\left(\frac{1}{\alpha}-1\right)}(\alpha-1)\alpha \mathcal{C}_{\alpha}(U),
    & 1<\alpha\le 2.
  \end{cases}
\end{equation}
(2) When $\alpha = \frac12$, $C_{\alpha}(\rho)$ reduces to $2
C_s(\rho)$, and we have
\begin{equation}\label{eq25}
\text{Cost}(U)
\geq\frac{1}{2d^{2}} \mathcal{C}_{s}(U).
\end{equation}
(3) When $\alpha\to 1$, for any fixed $\delta \in (0,1)$ and any state $\rho$ satisfying $p_{\min}\ge \delta$, one obtains $\mathrm{Cost}(U) \ge \frac{\ln 2}{\ln(1/\delta)}\,
\bigl|C_r(U\rho U^\dagger) - C_r(\rho)\bigr|$.
We define the $\delta$-restricted coherence power as
$C_r^{(\delta)}(U) := \max_{\rho:\,p_{\min}\ge \delta}
\bigl|C_r(U\rho U^\dagger) - C_r(\rho)\bigr|$.
Then
\begin{equation}\label{eq26}
\mathrm{Cost}(U) \ge -\frac{\ln2}{\ln\delta} \;\mathcal{C}_r^{(\delta)}(U).
\end{equation}
(4) For $\alpha\in (0,1)$, $\mathrm{Tr}(\rho^{\alpha})$ reaches its
maximum at the maximally mixed state $\rho = \mathbf{I}/d^{n}$,
where $\mathbf{I}$ denotes the identity operator, from which
$\mathrm{Tr}(\rho^{\alpha}) = d^{n(1-\alpha)}$. Then we have
\begin{equation}\label{eq27}
    \mathrm{Cost}(U) \;\geq\;
    d^{n(\alpha-1)} (1-\alpha)\alpha \, \mathcal{C}_{\alpha}(U).
\end{equation}
Moreover, if $n<\frac{2}{\alpha}$, then
$d^{n(\alpha-1)}>d^{\frac{2}{\alpha}(\alpha-1)}$, and therefore the
lower bound in Eq. (\ref{eq27}) is tighter than the one in Eq.
(\ref{eq22}).

It is observed that for pure input states $\rho$,
the resulting lower bounds of the circuit cost are independent of
$k$, and in particular, for $\alpha \in (0,1)$, the bounds are
independent of $k$, $n$ and $d$. Interestingly, choosing $d=2$ which
corresponds to the setting of an $n$-qubit system, the lower bound
in Eq. (\ref{eq25}) is $\frac{1}{8}\mathcal{C}_S(U)$, while the one
derived in Theorem 47 of \cite{Bu2024} is
$\frac{1}{8}\mathcal{C}_r(U)$. In the limit $\alpha \to 1$, the
lower bound in Eq. (\ref{eq26}) depend on $\delta$ if $p_{\min}\ge
\delta$. In particular, if $\delta>d^{-8}$, from Eq.
(\ref{eq26}) we have $\mathrm{Cost}(U) \ge \frac{1}{8\log
d}\mathcal{C}_r^{(\delta)}(U)$. This indicates that our lower bound
might be tighter than the one derived in \cite{Bu2024} for
appropriate chosen $\delta$.

Based on the definitions of $\mathrm{CGP}_{S}$ and $\mathrm{CGP}_{R}$, and using \cite{Bu2024} together with Eq.~(\ref{eq10}), we can derive the following corollary.\\
{\bf Corollary 1} The relationships between the circuit cost of a quantum circuit $U \in \mathrm{SU}(d^n)$ and the CGP defined respectively in terms of skew information and relative entropy are
\begin{equation}\label{eq28}
\text{Cost}(U)\geq\frac{1}{2d^{2}}\mathrm{CGP}_{S}(U)~~~ \mathrm{and} ~~~\text{Cost}(U)\geq\frac{1}{8\log d}\mathrm{CGP}_{R}(U).
\end{equation}
\indent Theorem~2 indicates that the Tsallis relative
$\alpha$-entropy of the cohering power can serve as a lower-bound
estimate for $\mathrm{Cost}(U)$. Although this bound can be
theoretically established, it is generally difficult to evaluate in
practice. In contrast, $\mathrm{CGP}$ given in Corollary~1 admit
explicit analytical expressions and are therefore computationally
tractable. Consequently, we can obtain a concrete lower-bound
estimate for $\mathrm{Cost}(U)$.

To illustrate this, Table~1 presents several examples, partially
adapted from \cite{Zhang2018,Wu2022}. In particular, we consider the
unitary operators $U_{\theta} = \left(\begin{smallmatrix}
\cos\theta & \sin\theta \\
-\sin\theta & \cos\theta
\end{smallmatrix}\right)$
and
$U_{t} =
\left(\begin{smallmatrix}
\sqrt{t}+\sqrt{1-t}\,\mathrm{i} & 0 & 0 & 0 \\
0 & \sqrt{t} & \sqrt{1-t}\,\mathrm{i} & 0 \\
0 & \sqrt{1-t}\,\mathrm{i} & \sqrt{t} & 0 \\
0 & 0 & 0 & \sqrt{t}+\sqrt{1-t}\,\mathrm{i}
\end{smallmatrix}\right)$, where $\theta \in [0,\pi]$ and $t \in [0,1]$. The lower bounds of $\mathrm{Cost}(U)$ for $U_{\theta}$ and $U_{t}$ are shown in Fig. 2.
\begin{table}[htbp]
\caption{The coherence generating power and lower bounds of
$\mathrm{Cost}(U)$ for typical quantum gates} \centering
\renewcommand{\arraystretch}{2.5}

\resizebox{1.0\textwidth}{!}{   % ←← 只改这里的 0.85 即可调整整体大小
\begin{tabular}{l c c c}
\hline\hline
Quantum Gate $U$
& $\mathrm{CGP}_S(U)$
& $\mathrm{CGP}_R(U)$
& The lower bound of $\mathrm{Cost}(U)$ \\
\hline

Hadamard ($H$)
& $\displaystyle \frac12\left(1-\frac{3\pi}{16}\right)$
& $\displaystyle \ln 2 - \frac12$
& $\displaystyle \frac{1}{16\ln 2}\left(1-\frac{3\pi}{16}\right)$ \\

$\sqrt{\text{swap}}$
& $\displaystyle \frac14\left(1-\frac{54545\pi}{262144}\right)$
& $\displaystyle \frac12\ln 2$
& $\displaystyle \frac{1}{16}$ \\

$U_\theta$
& $\displaystyle \frac12\left(1-\frac{3\pi}{16}\right)\sin^2(2\theta)$
& $\displaystyle \frac{\sin^4\theta\ln(\sin^2\theta)-\cos^4\theta\ln(\cos^2\theta)}{\cos2\theta}$
& $\displaystyle f(\theta)$ \\

$U_t$
& $\displaystyle t(1-t)\left(1-\frac{54545\pi}{262144}\right)$
& $\displaystyle \frac{t^2\ln t - (1-t)^2\ln(1-t)}{2(1-2t)}$
& $\displaystyle \frac{t^2\ln t - (1-t)^2\ln(1-t)}{16\ln 2(1-2t)}$ \\

CNOT, Toffoli, X, Y, Z, T
& 0 & 0 & 0 \\

\hline\hline
\end{tabular}
}

\end{table}

\begin{figure}[ht]\centering
\subfigure[] {\begin{minipage}[figure2a]{0.45\linewidth}
\includegraphics[width=1.0\textwidth]{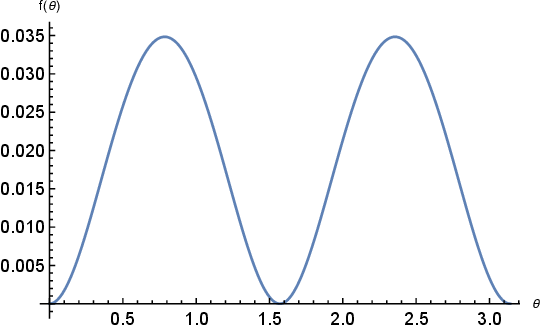}
\end{minipage}}
\subfigure[] {\begin{minipage}[figure2b]{0.45\linewidth}
\includegraphics[width=1.0\textwidth]{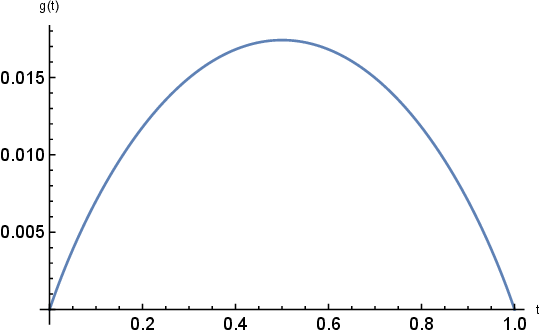}
\end{minipage}}
\caption{{The lower bounds of $\mathrm{Cost}(U)$. (a)
$U=U_{\theta}$, where the lower bound is denoted by $f(\theta)=
\max\left(
\frac{1}{16}\left(1-\frac{3\pi}{16}\right)\sin^{2}(2\theta),\;
\frac{\sin^{4}\theta\,\ln(\sin^{2}\theta)-\cos^{4}\theta\,\ln(\cos^{2}\theta)}{8\ln 2\cos 2\theta}
\right)$; (b)
$U=U_{t}$, where the lower bound is denoted by
$g(t)=\frac{t^2\ln t - (1-t)^2\ln(1-t)}{16\ln 2(1-2t)}$.\label{fig:Fig2}}}
\end{figure}
In the specific case of $d=2$ and
$n=5$ (i.e., $d^n=32$), the quantum Fourier transform (QFT) attains
$\mathrm{CGP}_S(F)\approx 0.2791$ and $\mathrm{CGP}_R(F)\approx
0.5875$. Substituting these quantities into the corresponding lower
bound inequality yields $\mathrm{Cost}(F)\ge 0.07343$. For the
Grover iteration in Grover's search algorithm, we obtain
$\mathrm{CGP}_S(G)\approx 0.0654$ and $\mathrm{CGP}_R(G)\approx
0.1487$, which in turn imply the lower bound $\mathrm{Cost}(G)\ge
0.0185$.

\vskip0.2in

\noindent {\bf 4. Imaginarity and circuit complexity}\\\hspace*{\fill}\\
In this section,  we investigate Tsallis relative $ \alpha $ entropy
of imaginarity and relative entropy of imaginarity in circuit
complexity, and derive lower bounds on the circuit cost based on
these imaginarity measures.

We define the Tsallis relative $ \alpha $ entropy of imaginaring power and the relative entropy of imaginaring power associated with a unitary evolution $U$ as
\begin{equation}\label{eq29}
\mathcal{M}_\alpha(U)= \max_{\rho \in \mathcal{D}((\mathbb{C}^d)^{\otimes n})} \left| M_\alpha\!\left(U \rho U^\dagger\right) - M_\alpha(\rho) \right|,
\end{equation}
and
\begin{equation}\label{eq30}
\mathcal{M}_r(U)= \max_{\rho \in \mathcal{D}((\mathbb{C}^d)^{\otimes n})} \left| M_r\!\left(U \rho U^\dagger\right) - M_r(\rho) \right|,
\end{equation}
where $\alpha \in (0,1)$ and the maximization is taken over all density operators $\rho$.
Building on this notion, we introduce the rate of imaginarity. For an $n$ qudit system initially prepared in state $\rho$, the rate of imaginarity based on Tsallis relative $ \alpha $ entropy and relative entropy are given by
\begin{equation}\label{eq31}
R^{\alpha}_{M}(H, \rho) := \frac{\mathrm{d}}{\mathrm{d}t} \, M_\alpha \!\left( e^{-\mathrm{i} t H} \rho \, e^{\mathrm{i} t H} \right) \bigg|_{t=0},
\end{equation}
and
\begin{equation}\label{eq32}
R_{M_r}(H, \rho) := \frac{\mathrm{d}}{\mathrm{d}t} \, M_r \!\left( e^{-\mathrm{i} t H} \rho \, e^{\mathrm{i} t H} \right) \bigg|_{t=0}.
\end{equation}
{\bf Theorem 3} Given a Hamiltonian $H$ on an $n$ qudit system and a quantum state $\rho \in \mathcal{D}\big((\mathbb{C}^d)^{\otimes n}\big)$,
the imaginarity rate based on Tsallis relative $ \alpha $ entropy satisfies the following bound
\begin{equation}\label{eq33}
|R_M^\alpha(H,\rho)| \le 4 \|H\|_\infty,
\end{equation}
where $\alpha \in (0,1)$.\\
$\it{Proof}$. Let $\rho_t = e^{-\mathrm{i} t H} \rho e^{\mathrm{i} t H}$. By the definition of the imaginarity rate, we have
\begin{align*}
R^{\alpha}_M(H, \rho)
&= \left. \frac{\mathrm{d}}{\mathrm{d}t} M_{\alpha}(\rho_t) \right|_{t=0}=\frac{\mathrm{d}}{\mathrm{d}t}\left[1 - \mathrm{Tr}\left(\rho_t^{\alpha} (\rho_t^{*})^{1-\alpha}\right)\right]\bigg|_{t=0}\\
&=-\sum_{j}\left\langle j\left|\frac{\mathrm{d}}{\mathrm{d}t}\rho_t^{\alpha}\cdot(\rho_t^{*})^{1-\alpha}+\rho_t^{\alpha}\cdot
\frac{\mathrm{d}}{\mathrm{d}t}(\rho_t^{*})^{1-\alpha}\right|j\right\rangle \bigg|_{t=0},
\end{align*}
where $\alpha \in (0,1)$. Direct calculations show that
$\frac{\mathrm{d}}{\mathrm{d}t}
\rho_t^{\alpha}\bigg|_{t=0}=\mathrm{i}\left[\rho^{\alpha},H\right]$,
and
\begin{align*}
\frac{\mathrm{d}}{\mathrm{d}t} (\rho_t^{*})^{1-\alpha}\bigg|_{t=0}
=\frac{\mathrm{d}}{\mathrm{d}t} \left(  e^{\mathrm{i} H^{*} t} \, (\rho^{*})^{1-\alpha} \, e^{-\mathrm{i} H^{*} t} \right)\bigg|_{t=0}
=\mathrm{i}\left[H^{*},(\rho^{*})^{1-\alpha}\right].
\end{align*}
By substituting the expressions, we have
\begin{align*}
R^{\alpha}_M(H, \rho)
&=-\mathrm{i}\sum_{j}\left\langle j\left|\left[\rho^{\alpha},H\right]\cdot(\rho^{*})^{1-\alpha}+\rho^{\alpha}\cdot
\left[H^{*},(\rho^{*})^{1-\alpha}\right]\right|j\right\rangle\\
&=-\mathrm{i}\mathrm{Tr}\left(\left[\rho^{\alpha},H\right]\cdot(\rho^{*})^{1-\alpha}+\rho^{\alpha}\cdot
\left[H^{*},(\rho^{*})^{1-\alpha}\right]\right)\\
&=\mathrm{i}\mathrm{Tr}\left(\left[\rho^{\alpha},(\rho^{*})^{1-\alpha}\right]\left(H+H^{*}\right)\right).
\end{align*}
According to the H\"older's inequality, we have
\begin{align*}
|R^{\alpha}_M(H, \rho)|
\leq\left\|\left[\rho^{\alpha},(\rho^{*})^{1-\alpha}\right]\right\|_{1}\|\left(H+H^{*}\right)\|_{\infty}.
\end{align*}
By the definition of the trace norm and the triangle inequality, one has
\begin{equation*}
\left\|\left[\rho^{\alpha},(\rho^{*})^{1-\alpha}\right]\right\|_{1}=\bigl\| \rho^{\alpha} (\rho^\ast)^{1-\alpha}
      - (\rho^\ast)^{1-\alpha} \rho^{\alpha} \bigr\|_1
\le
\bigl\| \rho^{\alpha} (\rho^\ast)^{1-\alpha} \bigr\|_1
+
\bigl\| (\rho^\ast)^{1-\alpha} \rho^{\alpha} \bigr\|_1 .
\end{equation*}
Since $\|\rho^{\alpha}\|_{1/\alpha}= \left( \operatorname{Tr} \rho \right)^{\alpha}
= 1$ and $\| (\rho^\ast)^{1-\alpha}\|_{1/(1-\alpha)}= \left( \operatorname{Tr} \rho^\ast \right)^{1-\alpha}
= 1$, by applying Hölder's inequality for Schatten norms, we obtain
\begin{equation*}
\|\rho^{\alpha} (\rho^\ast)^{1-\alpha}\|_1 \le \|\rho^{\alpha}\|_{1/\alpha} \, \| (\rho^\ast)^{1-\alpha}\|_{1/(1-\alpha)} = 1 .
\end{equation*}
By similar arguments, one can verify that $\|(\rho^\ast)^{1-\alpha}
\rho^{\alpha}\|_1 \le 1$. Substituting these bounds into the
previous commutator estimate, we arrive at
\begin{equation*}
\left\|\left[\rho^{\alpha},(\rho^{*})^{1-\alpha}\right]\right\|_{1} \le 2 .
\end{equation*}
Inserting the above inequality into the earlier bound gives
\begin{equation*}
|R_M^{\alpha}(H,\rho)| \le 2 \| H + H^\ast \|_\infty .
\end{equation*}
Noting that $ \| H + H^\ast \|_\infty \le \|H\|_\infty +
\|H^\ast\|_\infty = 2 \|H\|_\infty$, we get
\begin{equation*}
|R_M^\alpha(H,\rho)| \le 4 \|H\|_\infty .
\end{equation*}
This completes the proof.\qed

It can be seen that the imaginarity rate, defined
via the Tsallis relative $\alpha$-entropy for an
$n$-qudit quantum system, can be upper bounded by the operator norm of the Hamiltonian only, which is completely independent of both the system dimension and the entropic parameter $\alpha$. Consequently, Theorem~3 provides a unified and dimension-free characterization of imaginarity dynamics, demonstrating that the maximal rate of imaginarity is fundamentally limited by the intrinsic energy scale of the system.\\
{\bf Theorem 4} The circuit cost of a quantum circuit $U \in \mathrm{SU}(d^n)$ is
lower bounded by the Tsallis relative $ \alpha $ entropy of imaginaring power as follows
\begin{equation}\label{eq34}
\mathrm{Cost}(U) \;\geq\; \frac{1}{2\kappa_{\max}}\,M_\alpha(U)\ge \frac{1}{4}\,M_\alpha(U),
\end{equation}
where $\kappa_{\max}=\max_j \|o_j+o_j^{\ast}\|_\infty \in(0,2]$.\\
$\it{Proof}$. Fix $N\in\mathbb{N}$ and set $H_t=
H\!\left(\frac{t}{N}\right), \Delta t = \frac{1}{N}, t=1,\ldots,N$.
Define the discrete evolution by $\rho_t = e^{-\mathrm{i}\Delta t
H_t}\,\rho_{t-1}\,e^{\mathrm{i}\Delta t H_t}.$ For $s\in[0,\Delta
t]$, denote the intermediate state along the $t$-th segment by $
\rho_{t-1}(s)= e^{-\mathrm{i}sH_t}\,\rho_{t-1}\,e^{\mathrm{i}sH_t}.
$ Then, by the fundamental theorem of calculus,
\begin{equation*}
M_\alpha(\rho_t)-M_\alpha(\rho_{t-1})
= \int_{0}^{\Delta t}\frac{\mathrm{d}}{\mathrm{d}s}M_\alpha\!\bigl(\rho_{t-1}(s)\bigr)\,\mathrm{d}s .
\end{equation*}
Using $|R_M^{\alpha}(H_t,\rho)| \le 2 \| H_t + H_t^\ast \|_\infty $,
we obtain
\begin{equation*}
\bigl|M_\alpha(\rho_t)-M_\alpha(\rho_{t-1})\bigr|
\le \int_{0}^{\Delta t} 2\|H_t+H_t^{\ast}\|_\infty\,\mathrm{d}s
= \frac{2}{N}\|H_t+H_t^{\ast}\|_\infty .
\end{equation*}
Summing the above inequality over $t=1,\ldots,N$ and applying the triangle inequality yields
\begin{equation*}
\bigl|M_\alpha(\rho_N)-M_\alpha(\rho_0)\bigr|
\le \frac{2}{N}\sum_{t=1}^{N}\|H_t+H_t^{\ast}\|_\infty .
\end{equation*}
Letting $N\to\infty$, we obtain
\begin{equation*}
\bigl|M_\alpha(U\rho U^\dagger)-M_\alpha(\rho)\bigr|
\le 2\int_{0}^{1}\|H(s)+H(s)^{\ast}\|_\infty\,\mathrm{d}s .
\end{equation*}
Since
$H(s)+H(s)^{\ast}=\sum_j r_j(s)\bigl(o_j+o_j^{\ast}\bigr),$
and therefore
\begin{equation*}
\|H(s)+H(s)^{\ast}\|_\infty
\le \sum_j |r_j(s)|\,\|o_j+o_j^{\ast}\|_\infty
\le \kappa_{\max}\sum_j |r_j(s)|,
\end{equation*}
where we define
$\kappa_{\max}=\max_j \|o_j+o_j^{\ast}\|_\infty \le 2 .$
Substituting this estimate into the integral bound, we conclude that for any initial state $\rho$,
\begin{equation*}
\bigl|M_\alpha(U\rho U^\dagger)-M_\alpha(\rho)\bigr|
\le 2\kappa_{\max}\int_{0}^{1}\sum_j |r_j(s)|\,\mathrm{d}s .
\end{equation*}
Taking the infimum over all $r_j$ turns the right-hand side into
$2\kappa_{\max}\mathrm{Cost}(U)$, and taking the supremum over
$\rho$ yields $ M_\alpha(U)\le 2\kappa_{\max}\mathrm{Cost}(U). $ If
$\kappa_{\max}\neq 0$, we immediately have
\begin{equation*}
\mathrm{Cost}(U)\ge \frac{1}{2\kappa_{\max}}\,M_\alpha(U).
\end{equation*}
If $\kappa_{\max}=0$, we have $\|H(s)+H(s)^*\|=0$ for any admissible
$H(s)=\sum_j r_j(s)o_j$. By Theorem~3, $R_M^\alpha(H(s),\rho)=0$ for
all $s$ and $\rho$, hence $M_\alpha(U)=0$ and the bound is trivial.
Moreover, since $\kappa_{\max}\le 2$, it follows that
\begin{equation*}
\mathrm{Cost}(U)\ge \frac{1}{4}\,M_\alpha(U),
\end{equation*}
which completes the proof.\qed

From Theorem~4, the circuit cost is lower bounded by the Tsallis relative $\alpha$-entropy of imaginaring power, which is completely independent of both the system dimension and the entropic parameter $\alpha$. As a result, Theorem~4 provides a dimension-independent characterization of the minimal resource cost required to implement a quantum circuit in terms of its capacity to generate imaginarity.\\
{\bf Lemma 3}\cite{Aud2014,Mar2016} Let $X$ and $Y$
be positive trace class operators such that $X \le Y$, with
$\mathrm{Tr}(X)\in (0,1]$ and $\mathrm{Tr}(Y) = 1$. In finite
dimensional Hilbert spaces, we have
\begin{equation}\label{eq35}
\left\|\left[X, \log Y\right]\right\|_{1} \le 2\, g(\mathrm{Tr}(X)),
\end{equation}
where $g(t) = -t\log t - (1 - t) \log (1-t)$ denotes the binary
entropy function (with log taken to be base 2).

Based on this lemma, the following conclusions are presented.\\
{\bf Theorem 5} Given a Hamiltonian $H$ on an $n$ qudit system and a quantum state $\rho \in \mathcal{D}\big((\mathbb{C}^d)^{\otimes n}\big)$,
the imaginarity rate based on relative entropy satisfies the following bound
\begin{equation}\label{eq36}
\big| R_{M_r}(H,\rho) \big| \leq \frac{2g\!\left(t_{\ast}\right)}{t_{\ast}} \|H\|_{\infty}\le 4\|H\|_{\infty},
\end{equation}
where $t_{\ast} = \sup\{t\in(0,1] : t\rho \le \Delta_{1}(\rho)\}\in[\frac{1}{2},1]$.\\
$\it{Proof}$. Let $\rho_t = e^{-\mathrm{i} t H} \rho e^{\mathrm{i} t H}$. By the definition of the imaginarity rate, we have
\begin{equation*}
R_{M_r}(H, \rho) = \left. \frac{\mathrm{d}}{\mathrm{d}t} \Big[ S(\Delta_{1}(\rho_t)) - S(\rho_t) \Big] \right|_{t=0}=\left. \frac{\mathrm{d}}{\mathrm{d}t}S(\Delta_{1}(\rho_t)) \right|_{t=0},
\end{equation*}
where $\Delta_{1}(\rho) = \tfrac{1}{2}(\rho + \rho^T)$ represents the real part of the state.
To begin with, we calculate the derivative of $\Delta_{1}(\rho_t)$. Since $\rho_t^T = \left( e^{-\mathrm{i} t H} \rho e^{\mathrm{i} t H} \right)^T = e^{\mathrm{i} t H^{T}} \rho^T e^{-\mathrm{i} t H^{T}}$, we obtain
\begin{equation*}
\frac{\mathrm{d} \rho_t^T}{\mathrm{d}t} = \mathrm{i} H^{T} e^{\mathrm{i} t H^{T}} \rho^T e^{-\mathrm{i} t H^{T}} - e^{\mathrm{i} t H^{T}} \rho^T e^{-\mathrm{i} t H^{T}} \mathrm{i} H^{T}
= \mathrm{i} [H^{T}, \rho_t^T].
\end{equation*}
Similarly, we have $\frac{\mathrm{d} \rho_t}{\mathrm{d}t}=-\mathrm{i} [H, \rho_t]. $
Consequently, the derivative of $\Delta_{1}(\rho_t)$ is given by
\begin{equation*}
\frac{\mathrm{d}}{\mathrm{d}t} \Delta_{1}(\rho_t)
= \frac{1}{2} \left( \frac{\mathrm{d} \rho_t}{\mathrm{d}t} + \frac{\mathrm{d} \rho_t^T}{\mathrm{d}t} \right)
= \frac{1}{2} \left( -\mathrm{i} [H, \rho_t] + \mathrm{i} [H^{T}, \rho_t^T] \right).
\end{equation*}
Furthermore, it can be shown that
\begin{align*}
\left.\frac{\mathrm{d}}{\mathrm{d}t} S(\Delta_{1}(\rho_t)) \right|_{t=0}
=&- \mathrm{Tr} \left( \tfrac{1}{2} \left( -\mathrm{i} [H, \rho] + \mathrm{i} [H^T, \rho^T] \right) \log \Delta_{1}(\rho) \right)\notag\\
=&\frac{\mathrm{i}}{2} \mathrm{Tr} \left( [H, \rho] \log \Delta_{1}(\rho) \right)
- \frac{\mathrm{i}}{2} \mathrm{Tr} \left( [H^T, \rho^T] \log \Delta_{1}(\rho) \right).
\end{align*}
By using the property of trace and the definition of commutators, we have
\begin{align*}
R_{M_r}(H, \rho)
=&\frac{\mathrm{i}}{2} \mathrm{Tr} \big( [\rho, \log \Delta_{1}(\rho)] H \big)
+\frac{\mathrm{i}}{2} \mathrm{Tr} \big( [\rho, \log (\Delta_{1}(\rho))^T] H \big)\notag\\
=& \mathrm{i} \mathrm{Tr} \big( [\rho, \log \Delta_{1}(\rho)] H \big).
\end{align*}
Applying H\"older's inequality, we obtain
\begin{equation*}
\bigl|R_{M_r}(H,\rho)\bigr|
\le \bigl\|[\rho,\log \Delta_{1}(\rho)]\bigr\|_{1}\,\|H\|_{\infty}.
\end{equation*}
Define
\begin{equation*}
t_{\ast} = \sup\{t\in(0,1] : t\rho \le \Delta_{1}(\rho)\}.
\end{equation*}
Then $t_{\ast}\rho \le \Delta_1(\rho)$. Taking $X=t_{\ast}\rho$ and
$Y=\Delta_{1}(\rho)$ in Lemma~3, we have
\begin{equation*}
\left\|\left[t_{\ast}\rho, \log \Delta_{1}(\rho)\right]\right\|_{1}
\le 2\,g\!\left(t_{\ast}\right).
\end{equation*}
Since $\left[t_{\ast}\rho, \log \Delta_{1}(\rho)\right]=t_{\ast}[\rho,\log \Delta_{1}(\rho)]$, it follows that $\bigl\|[\rho,\log \Delta_{1}(\rho)]\bigr\|_{1} \le \frac{2\,g\!\left(t_{\ast}\right)}{t_{\ast}}$.
Hence, we obtain the refined bound
\begin{equation*}
\left|R_{M_r}(H,\rho)\right| \le \frac{2g\!\left(t_{\ast}\right)}{t_{\ast}}\,\|H\|_{\infty}.
\end{equation*}
Moreover, it follows that $t_{\ast}\ge 1/2$ since $\rho/2 \le
\Delta_{1}(\rho)$. Consequently, substituting $t_{\ast}=1/2$ into the bound
yields
\begin{equation*}
\bigl|R_{M_r}(H,\rho)\bigr| \le 4\|H\|_{\infty}.
\end{equation*}
This completes the proof.\qed

Comparing Theorem~3 and Theorem~5, it can be seen
that the imaginarity rate induced by the Tsallis $\alpha$ relative
entropy and the relative entropy both admit an upper bound which is
explicitly related to $\|H\|_{\infty}$ up to a same constant factor
4, but we can get a sharper bound of the relative entropy of
imaginarity rate here.

Building on Theorem 5, we can further derive the following result.\\
{\bf Theorem 6} The circuit cost of a quantum circuit $U \in
\mathrm{SU}(d^n)$ is lower bounded by the relative entropy of
imaginaring power,
\begin{equation}\label{eq37}
    \mathrm{Cost}(U) \geq\; \frac{1}{4} \, \mathcal{M}_r(U).
\end{equation}
$\it{Proof}$. The argument follows the
discretization and telescoping steps in Theorem~2. For any
implementation specified by control functions $r_j(s)$, take a
Trotter discretization $U \approx \prod_{t=1}^N U_t$ and define
$\rho_t = U_t \rho_{t-1} U_t^\dagger$. As in Theorem~2, approximate
each $U_t$ by its Lie-Trotter product $U_t^{(l)}=\prod_{q=1}^{lm}
W_{t,q}$ (a product of $lm$ elementary exponentials), and write the
total change as a telescoping sum over $q$. Let
$U_t=\lim_{l\to\infty}U_t^{(l)}$ and
$\rho_t^{(l)}=U_t^{(l)}\rho_{t-1}(U_t^{(l)})^\dagger$. In finite
dimensional case, since both of the map $\Delta_1$ and the von
Neumann entropy $S(\cdot)$ are continuous with respect to the trace
norm, $M_r(\rho)=S(\Delta_1(\rho))-S(\rho)$ is trace-norm
continuous. Since $U_t^{(l)} \to U_t(l\to \infty)$ in operator norm,
by similar arguments in Theorem 2, we have
$\|\rho_t^{(l)}-\rho_t\|_1 \to 0(l\to \infty)$. By the continuity of
$M_r(\rho)$, we get $M_r(\rho_t^{(l)})\to M_r(\rho_t)(l\to \infty)$,
and thus
\begin{equation*}
|M_r(\rho_t)-M_r(\rho_{t-1})| = \lim_{l\to\infty}
|M_r(\rho_t^{(l)})-M_r(\rho_{t-1})|.
\end{equation*}
Consequently, one may first bound
$|M_r(\rho_t^{(l)})-M_r(\rho_{t-1})|$ via the telescoping sum over
the $lm$ elementary factors. For each fixed $l$, the telescoping
argument yields the following single-step bound (which is
independent of $l$); hence the same bound also holds after taking
the limit $l\to\infty$.

For each elementary factor $\exp(-\mathrm{i}\tau
o_j)$, applying the fundamental theorem of calculus together with
the uniform form of Theorem~5,
\begin{equation*}
|R_{M_r}(H,\cdot)| \le 4\|H\|_\infty \qquad (\text{hence } |R_{M_r}(o_j,\cdot)| \le 4),
\end{equation*}
yields the single-step bound
\begin{equation*}
\big|M_r(\rho_t) - M_r(\rho_{t-1})\big|
\le \frac{4}{N}\sum_{j=1}^m \left|r_j\!\left(\frac{t}{N}\right)\right|.
\end{equation*}
Summing over $t$ and letting $N\to\infty$, we obtain for any $\rho$,
\begin{equation*}
\big|M_r(U\rho U^\dagger) - M_r(\rho)\big|
\le 4\int_0^1 \sum_{j=1}^m |r_j(s)|\,ds.
\end{equation*}
Taking the maximum over $\rho$ gives $M_r(U) \le 4\int_0^1
\sum_{j=1}^m |r_j(s)|\,ds,$ and finally taking the infimum over all
$r_j$ yields $\mathrm{Cost}(U) \ge \frac14 M_r(U)$. The proof is
complete.\qed

From Theorem~4 and Theorem~6, one can find that
the circuit cost can be lower bounded by the Tsallis $\alpha$
relative entropy of imaginaring power and the relative entropy of
imaginaring power up to a same constant factor 1/4, which are
independent of the system dimension. The two results characterize
the minimal resource to implement a quantum circuit in terms of its
capacity to generate imaginarity.

Many elementary gates that are ubiquitous in
quantum circuits, such as Pauli gates, CNOT gate, and Toffoli gate,
have zero imaginaring power with respect to a chosen reference
basis, meaning that they cannot create imaginarity from free inputs.
However, some other quantum gates give nonzero imaginaring power.
For example, for the $T$ gate, a real input state
$\rho=|+\rangle\langle+|$ provides explicit lower bounds
$\mathcal{M}_\alpha(T)\ge \tfrac12$ and $\mathcal{M}_r(T)\ge
0.6009$. Then we obtain $\mathrm{Cost}(T) \geq0.1502$, whose bound
is nontrivial. For the quantum Fourier transform $F$, choosing the
real input $\rho=\lvert 1\rangle\langle 1\rvert$ gives
$M_\alpha(F\rho F^\dagger)-M_\alpha(\rho)=1, M_r(F\rho
F^\dagger)-M_r(\rho)=1$. Hence, by definition, we have
$\mathcal{M}_\alpha(F)\ge 1$ and $\mathcal{M}_r(F)\ge 1$ (for $n\ge
2$). Therefore, we obtain $\mathrm{Cost}(F)\geq\frac{1}{4}$.
Nevertheless, many important quantum algorithms are typically
composed of a diverse set of gates and important components such as
QFT (e.g., Shor's algorithm, HHL algorithm, etc.) instead of a
single gate, the unitary corresponding to the circuit may have
nonzero imaginaring power. Consequently, lower bounds on circuit
cost derived from imaginaring power capture an intrinsic resource
requirement of the implementation and remain informative.

\vskip0.1in

\noindent {\bf 5. Conclusions and discussions}\\\hspace*{\fill}\\
In this study, we have explored the complexity of quantum circuits through the dual perspectives of quantum coherence and quantum imaginarity. We have also established the relationships between the circuit cost of a quantum circuit and the CGP defined respectively in terms of skew information and relative entropy. Based on Tsallis relative $\alpha$ entropy, we have established an upper bound on the coherence rate. Building upon this result, we have further derived a lower bound on the circuit cost via the Tsallis relative $\alpha$ entropy of cohering power, thereby uncovering the fundamental role of coherence in determining the resource requirements of quantum computation.

In addition, we have investigated circuit cost from the viewpoint of
quantum imaginarity, which has not been considered
in previous literatures to date. Utilizing Tsallis relative
$\alpha$ entropy and relative entropy, we have obtained upper bounds
for the imaginarity rate. Exploiting these properties, we have
subsequently derived corresponding lower bounds on the circuit cost
in terms of Tsallis relative $\alpha$ entropy of imaginaring power
and relative entropy of imaginaring power. In summary, these
findings provide new insights into the study of quantum circuit
complexity and contribute to a deeper understanding of the interplay
between coherence, imaginarity, and the resources required for
quantum information processing.

Note that the circuit cost can be lower bounded by
coherence/imaginarity implies that, if coherence/imaginarity grows
linearly with time, then the circuit cost must also grow linearly
with time, thereby offering insight into the short-time behavior of
complexity growth. Moreover, for quantum circuits, the imaginarity
power of individual quantum gates can be combined additively, under
appropriate composition rules, to yield tight bounds. The usefulness
of such bounds is clear: for example, they allow one to argue how
deep a weighted quantum circuit must be, at minimum, in order to
generate a prescribed coherence/imaginarity pattern in a desired
final state.

Coherence and imaginarity are both basis-dependent
resources, with incoherent states versus real states as free states,
and there are no analytical formulas for cohering power and
imaginaring power for arbitrary unitaries. Therefore, the lower
bounds of $\mathrm{Cost}(U)$ derived from them may capture different
facets of the ``nonclassical'' capability of a circuit with respect
to the chosen reference basis, and in general, we cannot clarify
which bound is tighter. Notably, certain commonly used gates may
yield zero lower bounds under cohering power (e.g.,
$\mathcal{C}_\alpha(T)=0$ for the $T$ gate), yet the bound is
nontrivial under imaginaring power (e.g., $\mathcal{M}_\alpha(T)\geq
\frac{1}{2}$ for the $T$ gate). On the other hand, choosing the real
input $\rho=\lvert 1\rangle\langle 1\rvert$ gives $C_\alpha(F\rho
F^\dagger)-C_\alpha(\rho)\approx 4.6569$ for $d=2$, $n=5$ and
$\alpha=2$. Hence, by definition, we have $\mathcal{C}_\alpha(F)\ge
4.6569$(for $n\ge 2$), which implies that $\mathrm{Cost}(F)\ge
0.8232$. This bound is tighter than the imaginarity-based bound
($\mathrm{Cost}(F)\geq\frac{1}{4}$). These observations suggest that
coherence and imaginarity bounds should be viewed as complementary
evidence for the circuit cost, and a robust strategy in applications
is to take the optimal one.

\vskip0.1in

\noindent

%=============================================================================%
\subsubsection*{Credit authorship contribution statement}
\small {Linlin Ye: Writing - original draft, Investigation, Conceptualization. Zhaoqi Wu: Writing - review and editing, Formal analysis, Methodology, Funding acquisition, Supervision. Nanrun Zhou: Writing - review and editing}

%=============================================================================%

%=============================================================================%
\subsubsection*{Declaration of competing interest}
\small {The author(s) declared no potential conflicts of interest with respect to the research, authorship, and/or publication of this article.}
%=============================================================================%

%=============================================================================%
\subsubsection*{Data availability}
\small {No new data were created or analysed in this study.}

%===========================================================================%

%=============================================================================%
\subsubsection*{Acknowledgements}
\small{The authors would like to thank Prof. Lin Zhang, Jianwei Xu
and Maosheng Li for helpful discussions. The authors would also like
to thank the referees for useful suggestions which greatly improved
the paper. This work was supported by National Natural Science
Foundation of China (Grant Nos. 12561084, 12161056); Natural Science
Foundation of Jiangxi Province (Grant No. 20232ACB211003).}

%===========================================================================%

\end{document}